\documentclass[preprint]{iucr}              
     \journalcode{S}              
\usepackage{xcolor}
\usepackage{enumitem}
\usepackage{amsmath,mathtools,calc}
\usepackage{url}
\usepackage{ulem}
\usepackage[utf8]{inputenc}
\usepackage{comment}
\usepackage{cleveref}

\graphicspath{{figs/}}

\newcommand{\supplementarysection}{%
  \setcounter{figure}{0}
  \let\oldthefigure\thefigure
  \renewcommand{\thefigure}{S\oldthefigure}
  \section{Supplementary section}
  \let\oldchapter\chapter
  \renewcommand{\chapter}{
    \let\thefigure\oldthefigure
    \let\chapter\oldchapter
    \oldchapter
  }
}

\usepackage{color,soul}
\usepackage[colorinlistoftodos]{todonotes}

\begin{document}                  



\title{Deep learning-based spatio-temporal fusion for high-fidelity ultra-high-speed x-ray radiography}



\cauthor[a]{Songyuan}{Tang}{tangs@anl.gov}{{address if different from \aff}}
\author[a,b]{Tekin}{Bicer}
\author[c]{Tao}{Sun}
\author[a]{Kamel}{Fezzaa}
\author[a]{Samuel J.}{Clark}

\aff[a]{Advanced Photon Source, Argonne National Laboratory, 60439 Lemont, IL, \country{USA}}
\aff[b]{Data Science and Learning Division, Argonne National Laboratory, 60439 Lemont, IL, \country{USA}}
\aff[c]{Department of Mechanical Engineering, Northwestern University, 60208 Evanston, IL, \country{USA}}









\maketitle                        

\begin{synopsis}
We develop and evaluate a deep learning-based algorithm that demonstrates the potential to reconstruct simultaneously high resolution, high frame-rate x-ray image sequences with high fidelity through spatio-temporal fusion. Our experimental evaluation shows that our method can significantly improve the accuracy of the reconstruction, achieving an average peak signal-to-noise ratio (PSNR) of more than 35 dB on two representative x-ray image sequences with input data streams of 4 times lower spatial resolution and 20 times lower frame rate, respectively.
\end{synopsis}

\begin{abstract}

Full-field ultra-high-speed (UHS) x-ray imaging experiments have been well established to characterize various processes and phenomena. However, the potential of UHS experiments through the joint acquisition of x-ray videos with distinct configurations has not been fully exploited. In this paper, we investigate the use of a deep learning-based spatio-temporal fusion (STF) framework to fuse two complementary sequences of x-ray images and reconstruct the target image sequence with high spatial resolution, high frame rate, and high fidelity. We applied a transfer learning strategy to train the model and compared the peak signal-to-noise ratio (PSNR), average absolute difference (AAD), and structural similarity (SSIM) of the proposed framework on two independent x-ray datasets with those obtained from a baseline deep learning model, a Bayesian fusion framework, and the bicubic interpolation method. The proposed framework outperformed the other methods with various configurations of the input frame separations and image noise levels. With 3 subsequent images from the low resolution (LR) sequence of a 4-time lower spatial resolution and another 2 images from the high resolution (HR) sequence of a 20-time lower frame rate, the proposed approach achieved an average PSNR of 37.57~dB and 35.15~dB, respectively. When coupled with the appropriate combination of high-speed cameras, the proposed approach will enhance the performance and therefore scientific value of the UHS x-ray imaging experiments. 
\end{abstract}

\section{Introduction}
Ultra-high-speed (UHS) cameras typically refer to cameras that record videos at a frame rate of above 1 MHz and exposure times of less than 1 µs \cite{tochigi2012global, sugawa2013ultra}. UHS cameras have been a key characterization tool utilized by a number of scientific user communities to investigate physical phenomena such as combustion, materials fracture, fluid dynamics, and electric discharges \cite{Miyauchi, Manin}. As modern sensor technologies continue to improve, UHS cameras operating at higher frame rates become commercially available \cite{Miyauchi, Nguyen, etoh2017theoretical}. 

In conventional full-field x-ray radiography, the concept of an indirect detector is a common experimental design choice which involves a scintillator to first convert the x-ray to the visible light, and a camera-lens system that subsequently collects the light to form the x-ray images \cite{bonse1996x,hartmann1975high,Koch}. Through the use of commercial cameras, indirect detectors greatly improve the flexibility of an x-ray experiment, leading to the development of customized instruments to meet the specific needs of a myriad of applications, such as additive manufacturing (and some other impactful use cases) \cite{Douissard}. When a high-speed camera is used in such experiments, however, a trade off generally exists between its spatial resolution and its temporal resolution \cite{Li}. In order to observe highly dynamic features at extremely fine temporal and spatial scales, imaging systems require frame rates and spatial resolutions that are not currently available. In other words, we face scientific demand for the highest acquisition frequency, largest field-of-view (FOV) and highest spatial resolution simultaneously, an experimental trilemma necessitating compromise for each experiment. As a result, in an experiment that normally utilizes a single high-speed camera implementation of different imaging settings, it is commonly required to repeat the same experiment multiple times to deploy the corresponding configuration. In the past, several multiplex setups consisting of more than one high-speed cameras have also been proposed to image the same experimental phenomena with various detector configurations \cite{luo2012gas, Ramos, Escauriza}. So far, there has been no study within the area of full-field UHS x-ray imaging, and limited studies in other related areas to fuse multiple high-speed acquisitions to aggregate distinct advantages of their underlying configurations. An exemplar configuration available at the 32-ID high-speed imaging beamline \cite{Parab} at the Advanced Photon Source (APS) would be to leverage simultaneously an UHS HPV-X2 (Shimadzu Corp., Japan) camera and a HS TMX7510 (Vision Research Inc., USA) camera which at full resolution have maximum frame rates of 5 MHz (400$\times$250 pixels) and 76 kHz (1280$\times$800 pixels), respectively.

Beyond x-ray imaging, \cite{Kornienko} developed a method to illuminate the object with three temporally resolvable nanosecond laser pulses during one single camera exposure and unmix the resulting image signals at the post-processing time \cite{Kornienko}. This method, termed “frequency recognition algorithm for multiple exposures” (FRAME), allowed efficient recording of complex processes with motion dynamics manifested at multiple time scales. \cite{He} designed a multi-modal acquisition framework to image ultrafast phenomena with high fidelity. In their work, three imaging models, namely, the compressed ultrafast photography (CUP), transient imaging at a relatively lower frame rate, and spatio-temporal integration were implemented within one single acquisition and fused with an untrained neural network (UNN). The reconstructed transient image sequence had improved frame rates, with demonstrated higher peak signal-to-noise ratio (PSNR) and structural similarity (SSIM) than prior methods.

To build general-purpose video processing software, various forms of object motion need to be properly handled \cite{xue2019video}. In particular, for the task of image restoration, the availability of neighboring frames, including those from co-registered imaging devices, could significantly extend the performance limit that can be achieved with a single image. One technique to restore low-quality video frames is called feature alignment. Broadly speaking, feature alignment methods can be categorized as explicit or implicit. Explicit methods first estimate and then compensate for motion fields between neighboring video frames to achieve alignment of corresponding features \cite{werlberger2011optical,fransens2007optical,caballero2017real}. Recent work on explicit feature alignment has concentrated on the use of deep learning to improve the accuracy of the estimated motion fields and better adapt to specific vision tasks \cite{wang2020deep}. Due to the powerful feature extraction capability of deep neural networks, and compounded with the efficient development of deformable convolutions \cite{dai2017deformable,zhu2019deformable}, implicit feature alignment has recently gained great momentum. The deformable convolution was originally designed to improve a neural network's adaptability in modeling varying object shapes and has demonstrated great success in applications such as object detection and semantic segmentation \cite{dai2017deformable}. Later, Tian et al. \cite{Tian} proposed a novel deep neural network, termed temporally deformable alignment network (TDAN) for video super-resolution. More specifically, the authors extended the use of deformable convolutions to dynamically predict offsets and align deep features between temporally neighboring image frames. Wang et al. \cite{Wang:19} further extended the deformable convolution module in TDAN to a hierarchical architecture to tackle larger object motions in videos of more dynamic scenes.

Despite that accurate and reproducible video fusion software has been developed, there is yet to have been uptake in routine UHS x-ray radiography experiments. The availability of such a capability could fundamentally transform current research activities to enable the discovery of new phenomena of scientific value, while simultaneously providing simplified, user-friendly experimental workflows. In other research areas such as communication network and remote sensing, spatio-temporal image fusion technologies have received a great amount of interest, with emerging computational model architectures, data formats, and machine learning algorithms demonstrating competitive performance \cite{Liu, Lu, Chen, Xiao}. Knowledge obtained in these developments could in turn be leveraged to improve UHS x-ray video fusion and reconstruction.

In this paper, we demonstrate the technical feasibility of fusing two sequences of high-speed x-ray images using a deep learning-based approach, with the intent of understanding the performance and limitations in reconstructing a single sequence of x-ray images with high spatial resolution and high frame rate. More specifically, we reorganized a video restoration framework with enhanced deformable convolution (EDVR) that fused a set of temporally consecutive low resolution (LR) image frames, so it could integrate details from additional high resolution (HR) input image frames to predict a target HR image frame with improved fidelity. This new model architecture aiming at spatio-temporal fusion, termed as EDVR-STF, was pre-trained on image restoration benchmarks and subsequently fine-tuned on dedicated x-ray image datasets for transfer learning. Performance was benchmarked against three existing approaches for image sequence restoration. In addition, we identified a new set of performance metrics, termed as the ``backward attention score'' and ``forward attention score'' in order to interpret the effectiveness of the proposed method and give insight into the utilization of the input HR frames, in the absence of the actual HR frames as the ground truth. 

\section{Methods and numerical experiments}\label{sec:num_exp}

\begin{figure}
    \label{fig:exp_illu}
    \centering
    \includegraphics[width=1\textwidth]{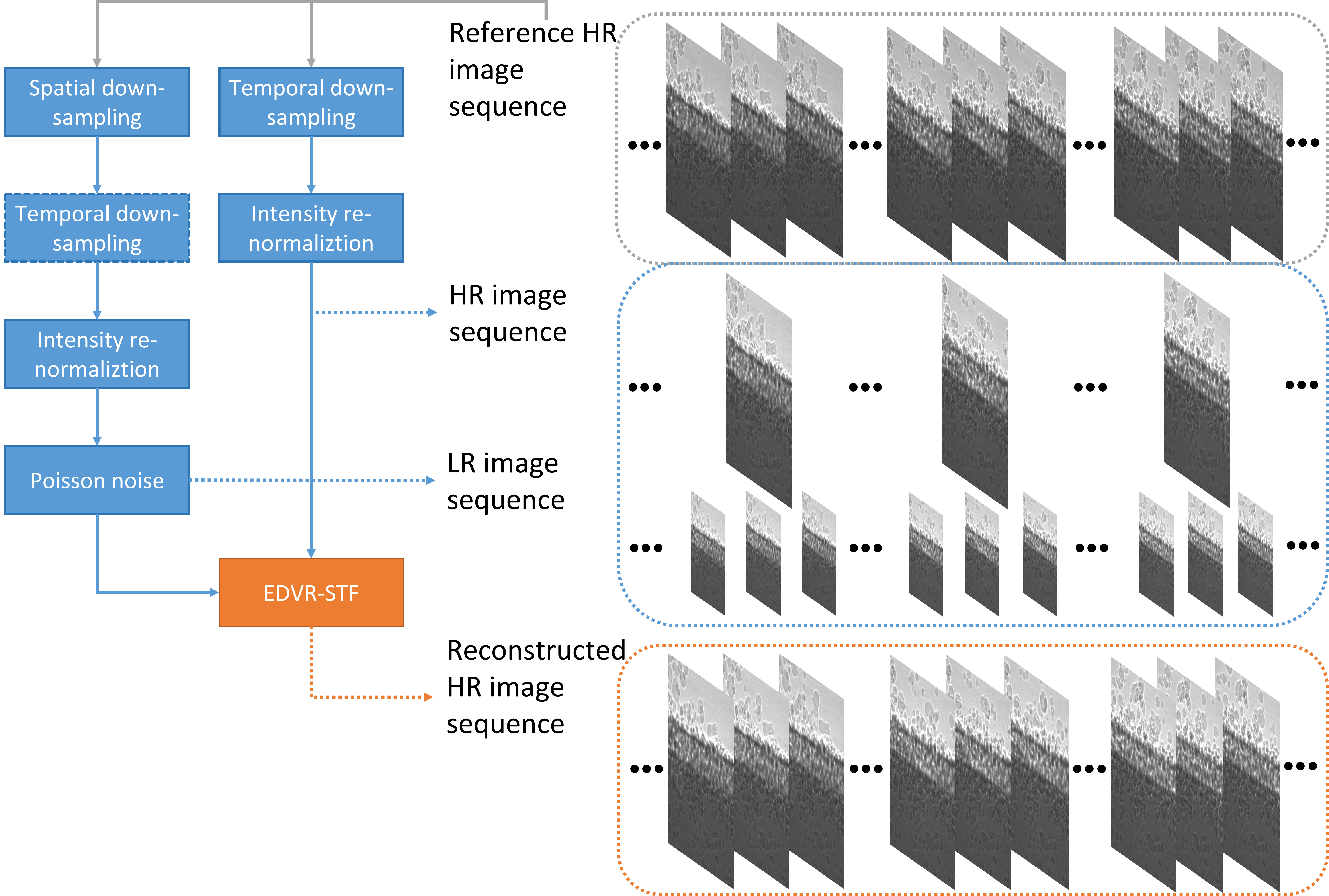}
    \caption{Illustration of the numerical experiment workflow.}
\end{figure}

The model architecture and model training were based on the BasicSR package \cite{basicsr}. More details can be found in Section \ref{sec:supp_method} of the supplementary information. 
The training data were obtained from high-speed X-ray imaging experiments conducted at the Argonne National Laboratory APS, on beamline 32 ID-B. The x-ray used for imaging was a pseudo-pink beam with a first harmonic energy of $\sim$ 24.5 keV and an energy bandwidth of $\sim 7\%$ generated with a 1.8 cm short period undulator with a gap of 12.5 mm and a sample-to-detector distance of typically $\sim$ 400 mm. A 100 µm thick LuAG:Ce scintillator was used to convert the x-ray to the light. The camera used was a Photron Fastcam SA-Z 2100K operating at a frame rate of 50 kHz, with an exposure time of 1 $\mu$s and a spatial resolution of 2 $\mu$m/pixel.
Two independent x-ray videos for applications of additive manufacturing and friction stir welding were used to test the performance of the proposed image reconstruction algorithm. Both videos were acquired using the same Photron camera as used in the acquisition of the training data, operated at frame rates of 50 kHz and 20 kHz, respectively at the 32-ID beamline of the APS \cite{Ren} and \cite{Agiwal}. For each type of video, the pixel values were normalized, and spatial and temporal sub-samples were obtained at varying sampling rates to create the LR (and with ultra-high speed) image sequence, and the HR (and with high speed) image sequence, respectively. In the case of the LR image sequence, Poisson noise of varying magnitudes was also simulated. More details refer to Section \ref{sec:supp_data_preproc} of the supplementary information. Model inference was thus performed by iterating through each LR frame as the reference LR frame of the input, pairing it with another two LR frames before and after it at the designated frame separation and with two nearest HR frames before and after it, from the designated down-sampled frame sequences, respectively, and outputting the reconstructed HR frame at the same time of the reference LR frame. Figure~\ref{fig:exp_illu} illustrates the experimental configuration.

In addition to the proposed EDVR-STF reconstruction model, the same testing data were also used to evaluate the bicubic interpolation, a Bayesian image fusion framework previously developed for geospatial applications \cite{xue2017bayesian}, and the baseline EDVR reconstruction model for video super-resolution \cite{Wang:19}. More details refer to Section \ref{sec:supp_baselines} of the supplementary information. To evaluate the candidate reconstruction algorithms, a total of 3 performance metrics were considered: the peak signal-to-noise ratio (PSNR) \cite{Wang:19}, the average absolute difference (AAD) \cite{Zhang}, and the structural similarity (SSIM) index \cite{Wang:04}. In the remainder of this paper, PSNR evaluated on the LR images with Poisson noise (relative to the LR images directly down-sampled from the original HR images) will be referred to as the ``LR image PSNR'', to be distinguished from that evaluated on the reconstructed HR images (relative to the original HR images). In addition to these reference-based quality assessment metrics, we also analyze the contribution of each of the input HR images to the reconstruction of the target HR image. More details refer to Section \ref{sec:supp_attention_scores} of the supplementary information.

\section{Results}
In this section, we report qualitative and quantitative analyses of the proposed EDVR-STF image reconstruction approach as applied to the two x-ray image sequences described in Section \ref{sec:num_exp} (paragraph 1) and Section \ref{sec:supp_data_preproc} of the supplementary information, i.e., in applications of additive manufacturing and friction stir welding, respectively. For the purpose of illustration, we first show temporally contiguous LR, HR, and reconstructed HR images of each dataset in one selected time window (with 20 consecutive frames) (\Cref{fig:5,fig:6,fig:7}) and in a longer time span (with 200-300 consecutive frames; supplementary videos), and compare local features before and after the image reconstruction. We then compare the proposed approach with the other 3 approaches described in Section \ref{sec:num_exp} (paragraph 2) and Section \ref{sec:supp_baselines} of the supplementary information, i.e., bicubic interpolation, Bayesian fusion, and EDVR super-resolution, when the LR images are subjected to the Poisson noise with varying levels of PSNR (Figure~\ref{fig:2}) and when the underlying frame rates of both LR and HR images vary (\Cref{fig:3,fig:S2,fig:S3}). To better interpret the behavior of the proposed deep learning-based approach, we show the attention scores of the target image to each corresponding HR image input to the algorithm as described in Section \ref{sec:supp_attention_scores} of the supplementary information and illustrate a decreasing pattern as the target image gets farther from the input HR image (\Cref{fig:4,fig:S4}). Last, we make a brief characterization of the computation times of each method under their most accessible implementations (Figure~\ref{fig:9}).

Figure~\ref{fig:5} illustrates typical results of the HR image reconstruction from 3 subsequent LR images (1 frame apart) and 2 fixed HR images (spaced by 20 LR frames), for the application of additive manufacturing (case 1). In particular, the results shown in C2-C4 are from time points when the actual HR images are not available (i.e., B2-B4 are “unseen” ground truth images held out at testing time), whereas C1 and C5 are from time points when the actual HR images are available (i.e., B1 and B5 as the preceding HR images in the reconstruction input). The normalized attention scores for C2, C3, and C4 are 0.69, 0.41, and 0.32 (backward attention) and 0.30, 0.39, and 0.70 (forward attention), respectively. Figure~\ref{fig:6} highlights moving features of the reconstructed image (in comparison with the original HR and LR images) in Figure~\ref{fig:5} with equal frame separation from both the preceding and succeeding HR images. When the motionless image content is masked out (by pixel-wise division with the preceding image as typical processing in the field of additive manufacturing), the reconstructed images show the moving particles and the keyhole features with significantly improved clarity compared with the LR image (A). The LR image sequence, HR image sequence, and the reconstructed HR image sequence in the longer time span are presented in the supplementary video.

\subsection{Qualitative assessments}
\begin{figure}
    \label{fig:5}
    \centering
    \includegraphics[width=0.8\textwidth]{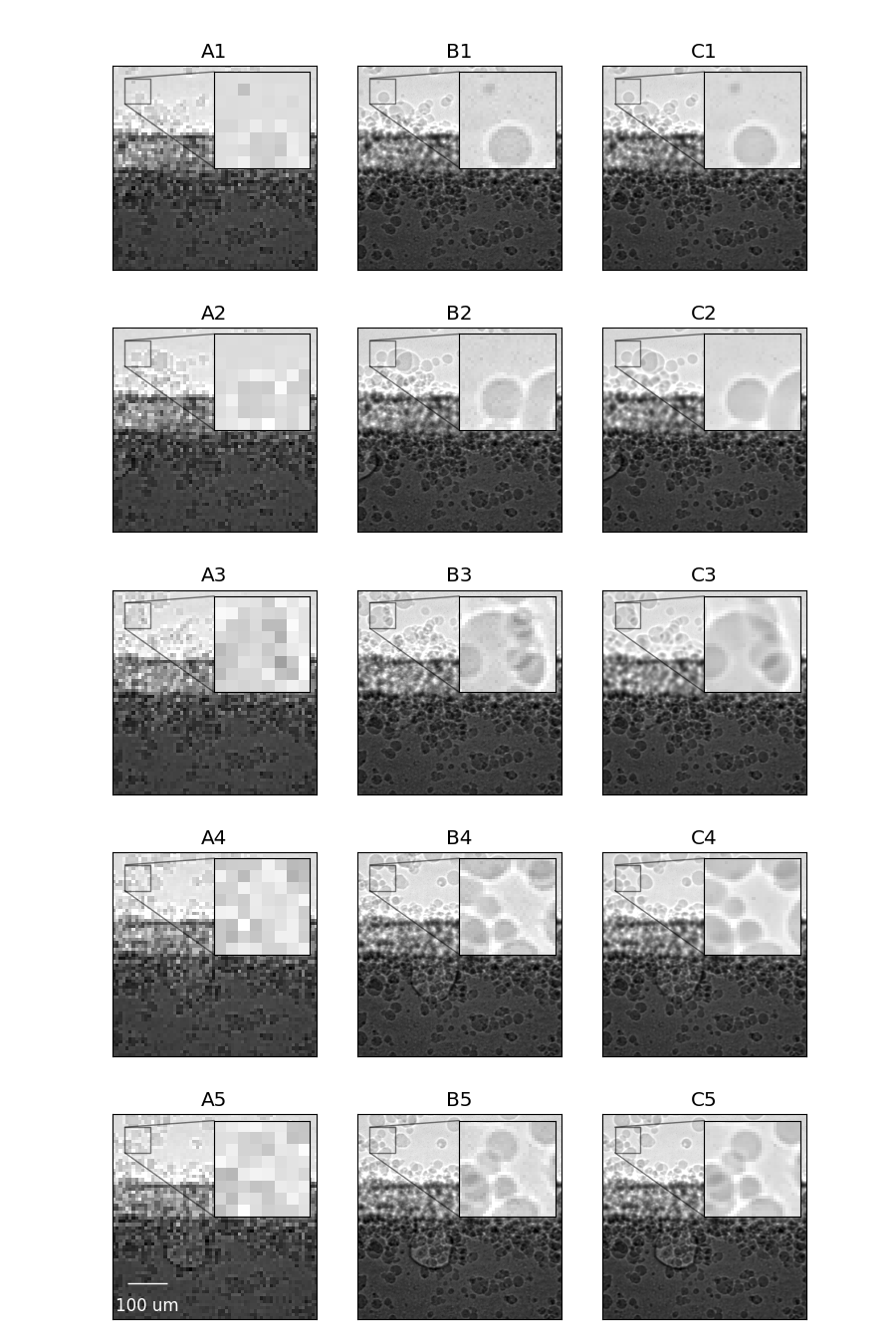}
    \caption{Selected image sequences of the LR images (column A), HR images (column B), and HR images reconstructed using EDVR-STF (column C) from case 1. For each image sequence, images are shown from frames i-10 (row 1), i-9 (row 2), i (row 3), i+9 (row 4), and i+10 (row 5), respectively. To reconstruct HR images in the frame range [i-10, i+10], only HR images from frames i-10 and i+10 were used. No Poisson noise was generated for the testing data. 
    }
\end{figure}

\begin{figure}
    \label{fig:6}
    \hspace*{-1.5cm}
    \includegraphics[width=1.2\textwidth]{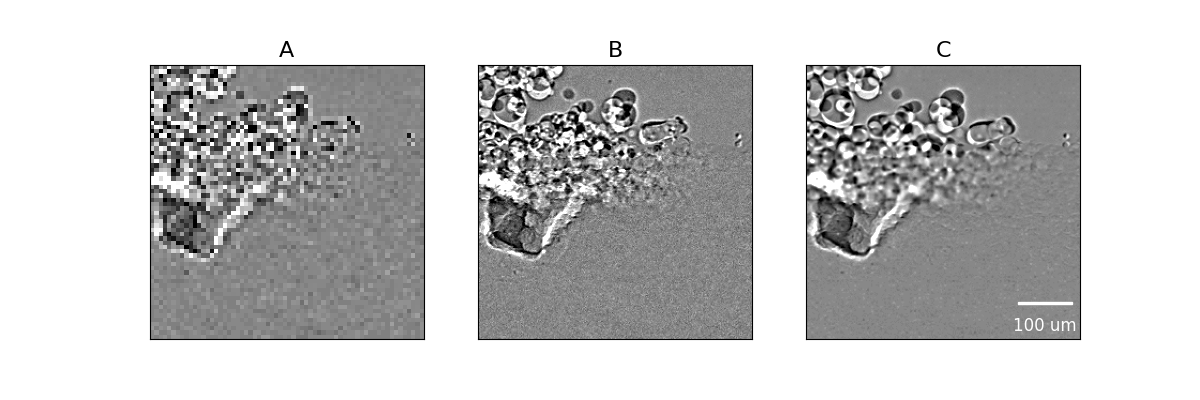}
    \caption{Selected image sequences of the LR image (A), HR image (B), and HR images reconstructed using EDVR-STF (C) from column C of Figure~\ref{fig:5}, divided, on the basis of the pixel value, by the same from their previous frame.}
\end{figure}

\begin{figure}
    \label{fig:7}
    \centering
    \includegraphics[width=0.8\textwidth]{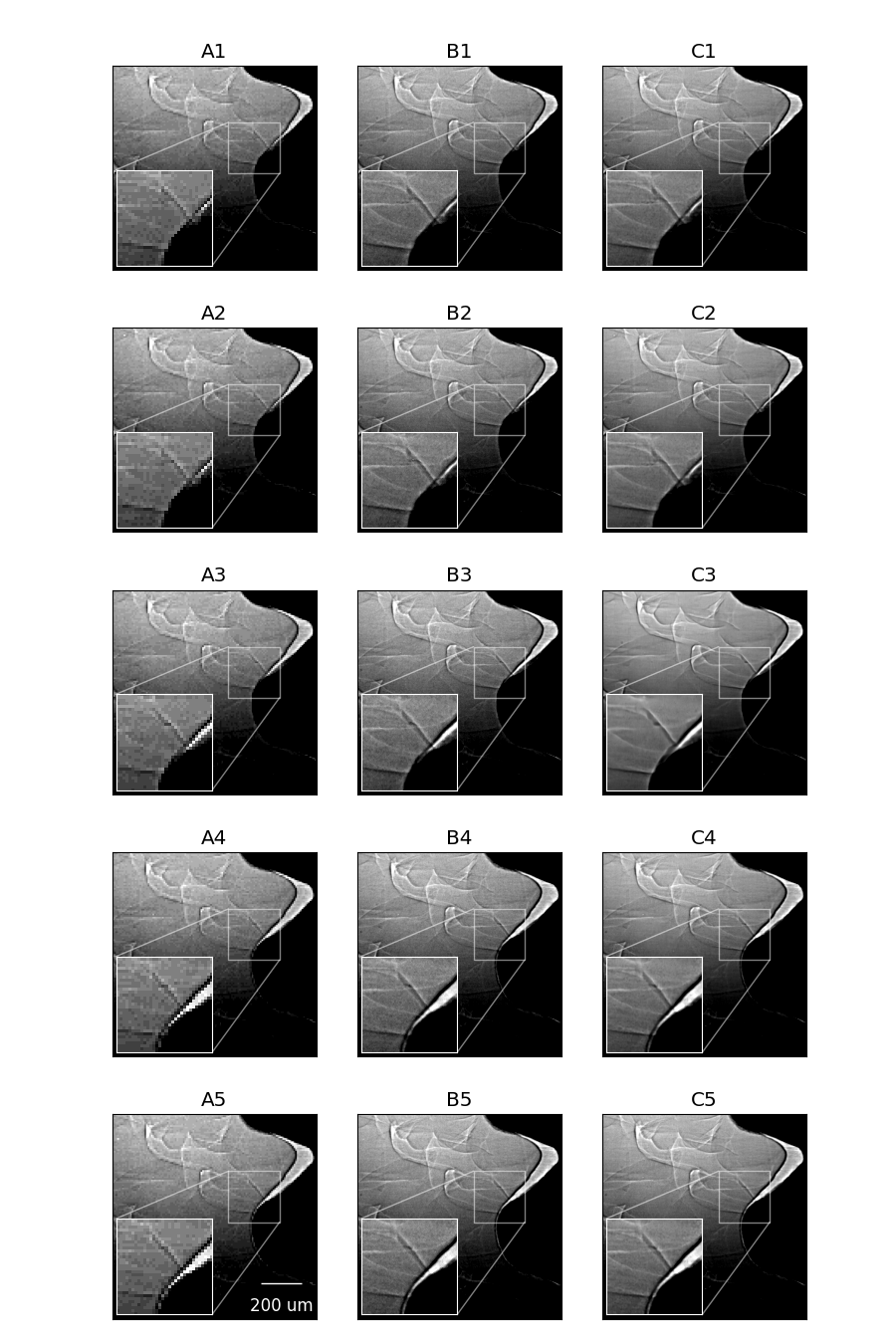}
    \caption{Selected image sequences of the LR images (column A), HR images (column B), and HR images reconstructed using EDVR-STF (column C) from case 2. For each image sequence, images are shown from frames i-10 (row 1), i-9 (row 2), i (row 3), i+9 (row 4), and i+10 (row 5), respectively. To reconstruct HR images in the frame range [i-10, i+10], only HR images from frames i-10 and i+10 were used. No Poisson noise was generated for the testing data.  
    }
\end{figure}

Figure~\ref{fig:7} shows another case of HR image reconstruction by applying the same model and pre/post-processing pipeline as in Figure~\ref{fig:5} to a distinct type of x-ray image sequence capturing the friction stir welding (case 2) \cite{Agiwal}. As can be seen from the images, the object motion captured in images from Figure~\ref{fig:7} is significantly slower than that from Figure~\ref{fig:5}. And the degradation in image quality after a 4 times down-sampling is visually less pronounced. The normalized attention scores for C2-C4 are 0.62, 0.36, and 0.34 (backward attention) and 0.34, 0.38, and 0.65 (forward attention) respectively. In particular, fine-scale details of the original LR, HR, and the reconstructed HR images in Figure~\ref{fig:7} are shown in the insets of the corresponding sub-figure. Overall, the reconstructed images show noticeable improvement in restoring image textures compared to the original LR image (A). The LR image sequence, HR image sequence, and the reconstructed HR image sequence in the longer time span are presented in the supplementary video.

\subsection{Performances with varying LR image PSNRs}\label{sec:poisson_perf}
\begin{figure}
    \label{fig:2}
    \hspace*{-1.5 cm}
    \vspace*{-1.5 cm}
    \includegraphics[width=1.2\textwidth]{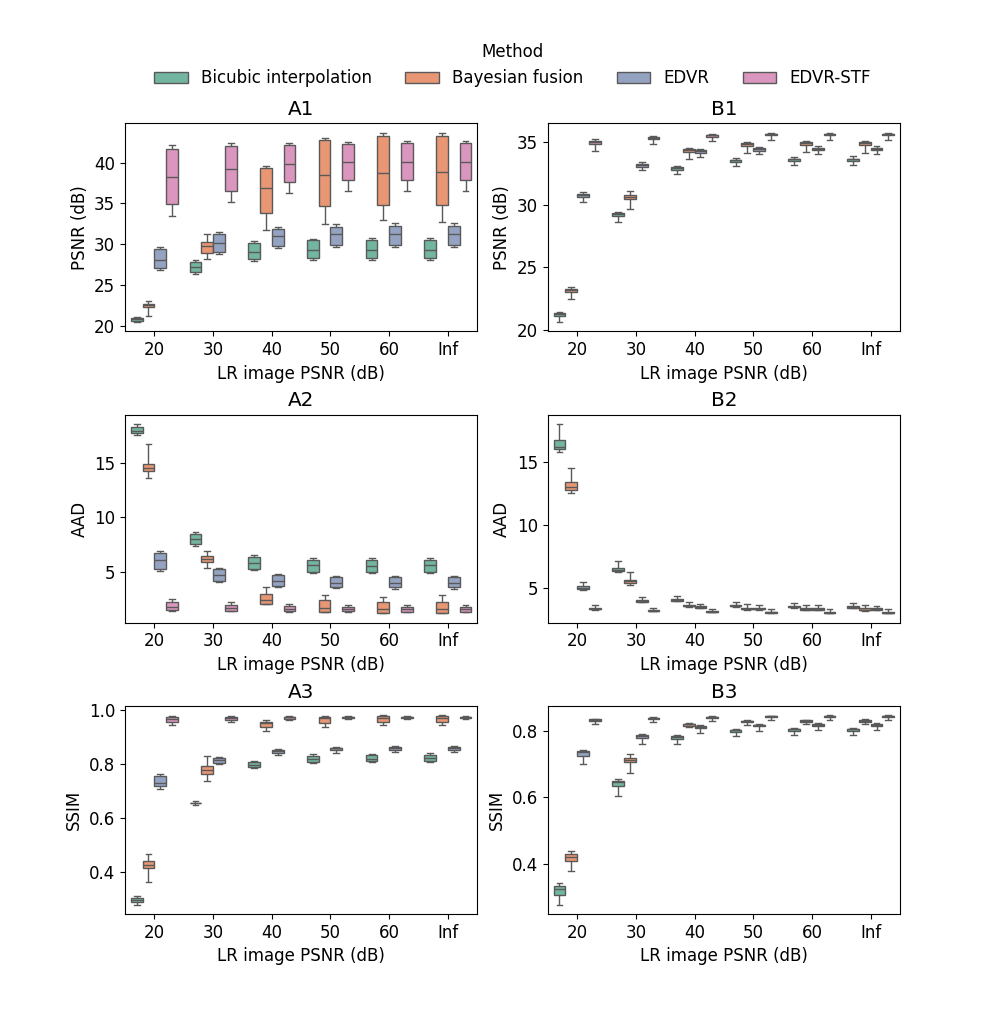}
    \caption{Reconstructed HR frame PSNR (dB) (row 1), AAD (row 2), and SSIM (row 3) each as a function of the LR image PSNR that was used to generate the Poisson noise in the input LR frames, based on bicubic interpolation, Bayesian fusion, EDVR, and EDVR-STF. Results were evaluated on case 1 (column A) and case 2 (column B) and presented as box plots. Under each testing condition, the same samples as used for EDVR-STF were used to test all other 3 algorithms.
    }
\end{figure}

Figure~\ref{fig:2} shows the PSNR, AAD, and SSIM of the 4 reconstruction algorithms evaluated on the 2 testing data sets with varying LR image PSNRs due to Poisson noise. Overall, due to data augmentation of the LR frames with Poisson noise during the training stage, both EDVR and EDVR-STF show more steady performance across different noise levels of the testing data. When the noise level is high (e.g., 20~dB LR image PSNR), deep learning-based algorithms perform significantly better than conventional approaches, with higher PSNR and SSIM, and lower AAD. As the LR image PSNR increases, the performance of the Bayesian fusion framework improves significantly, outperforming EDVR at the LR image PSNR of 40~dB and 50~dB in case 1 and case 2, respectively. At baseline with no Poisson noise generation in the testing data, the max PSNR of the Bayesian fusion framework is higher than EDVR-STF (47.03~dB vs 43.81~dB) in case 1 and lower than EDVR-STF (37.24~dB vs 39.41~dB) in case 2. The median PSNR of the Bayesian fusion framework is lower than that of EDVR-STF, consistently in both cases. At each noise level in both case 1 and case 2, the Bayesian fusion framework, EDVR, and EDVR-STF perform significantly better than bicubic interpolation. The other two performance indices show consistent patterns among the 4 methods and as the blank scan factor changes.

\subsection{Performances with varying image separations}
\begin{figure}
    \label{fig:3}
    \hspace*{-1.5 cm}
    \vspace*{-1.5 cm}
    \includegraphics[width=1.2\textwidth]{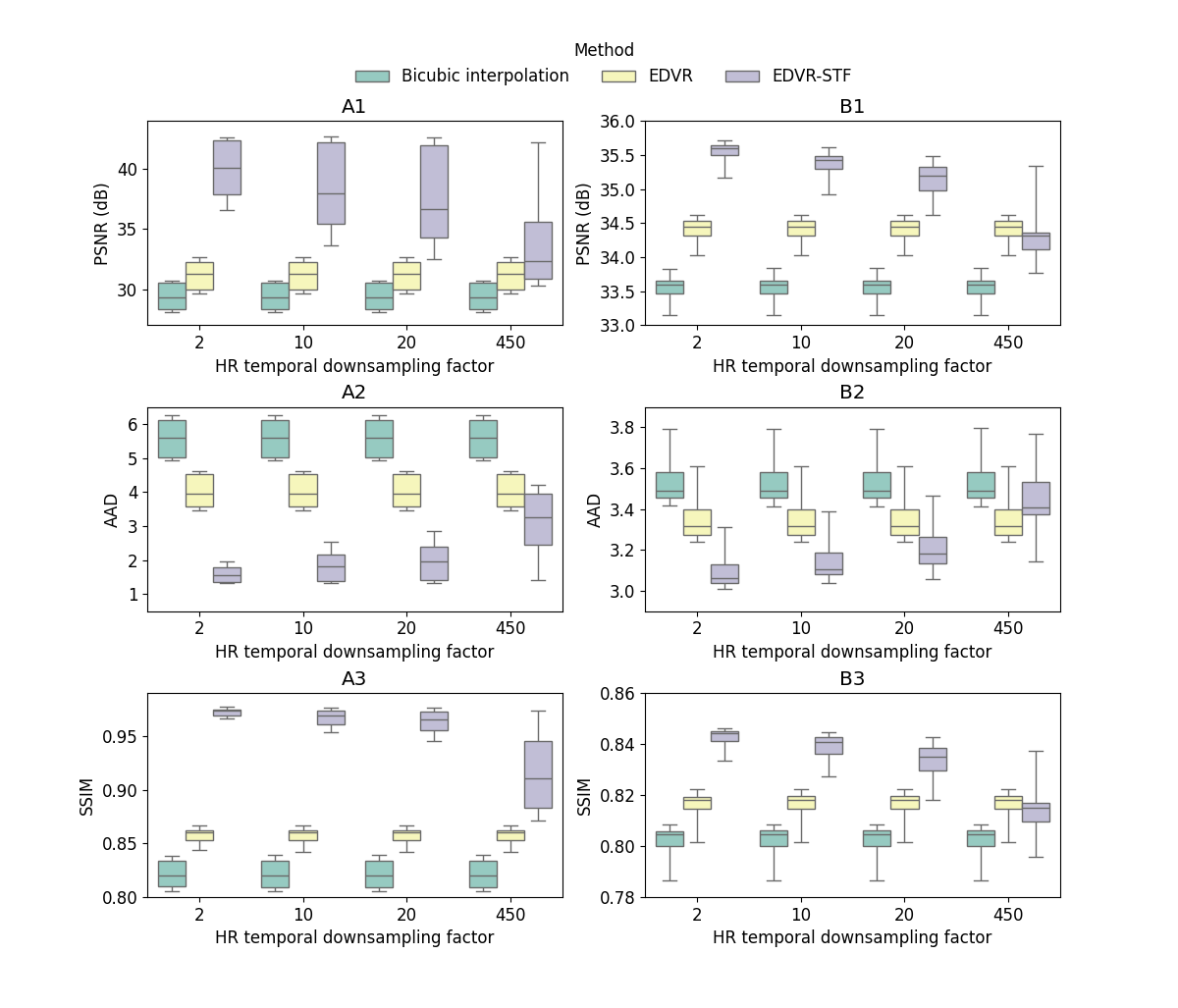}
    \caption{Reconstructed HR frame PSNR (dB) (row 1), AAD (row 2), and SSIM (row 3) each as a function of the HR frame sequence down-sampling factor, based on bicubic interpolation, EDVR, and EDVR-STF. Results were evaluated on case 1 (column A) and case 2 (column B) and presented as box plots. LR separation was 1 across all plots. Under each testing condition, the same samples as used for EDVR-STF were used to test all other 2 algorithms. No Poisson noise was generated for the testing data.}
\end{figure}

\begin{figure}
    \label{fig:S3}
    \hspace*{-1.5 cm}
    \vspace*{-1.5 cm}
    \includegraphics[width=1.2\textwidth]{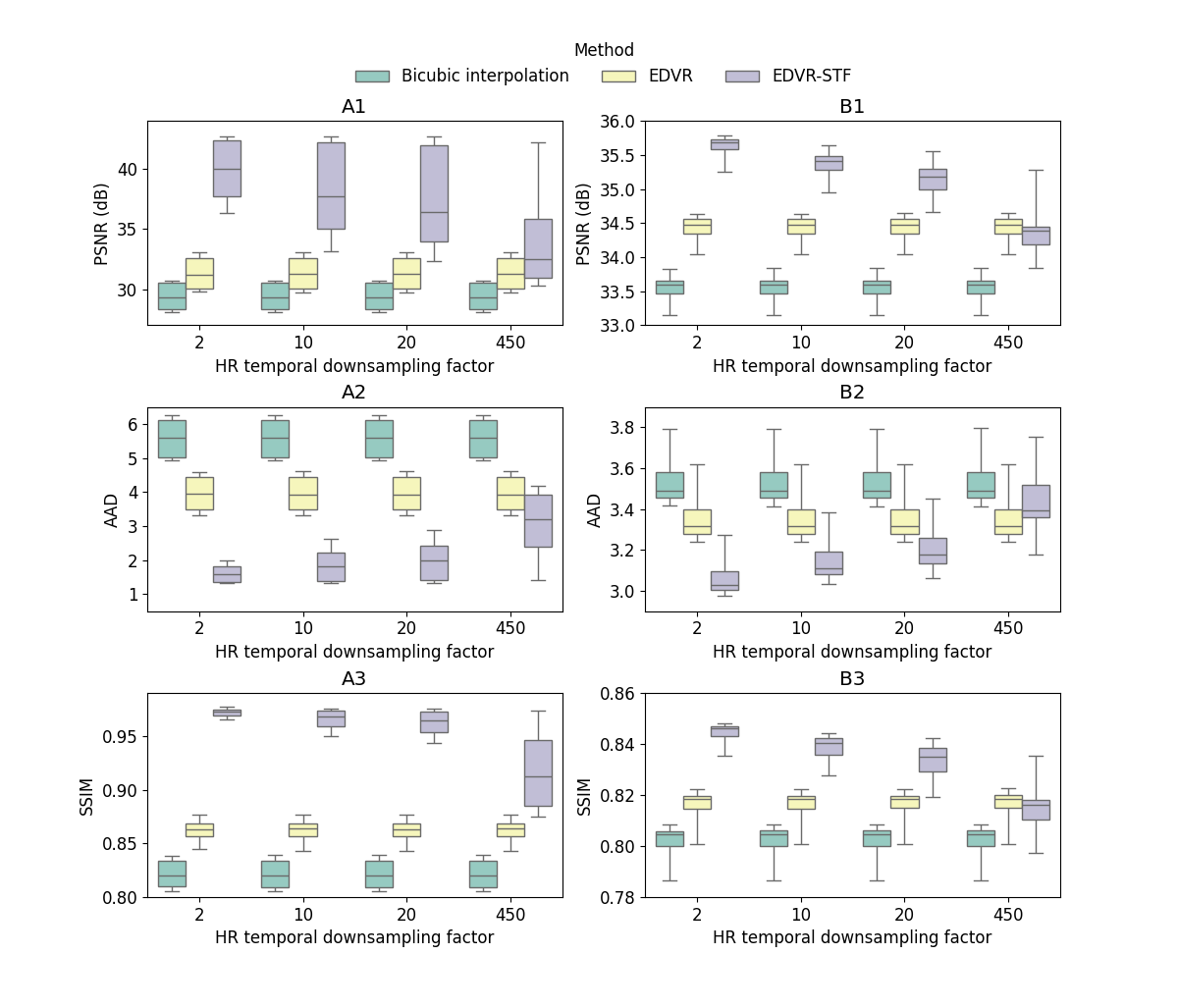}
    \caption{Reconstructed HR frame PSNR (dB) (row 1), AAD (row 2), and SSIM (row 3) each as a function of the HR frame sequence down-sampling factor, based on bicubic interpolation, EDVR, and EDVR-STF. Results were evaluated on case 1 (column A) and column 2 (column B) of the main paper and presented as box plots. LR separation was 3 across all plots. Under each testing condition, the same samples as used for EDVR-STF were used to test all other 2 algorithms. No Poisson noise was generated for the testing data.}
\end{figure}

Figure~\ref{fig:3} shows the PSNR, AAD, and SSIM of bicubic interpolation, EDVR, and EDVR-STF evaluated on the 2 testing data sets with varying down-sampling factors to temporally down-sample the HR image sequence and with the original frame separation of the LR image sequence. Overall, the performance of EDVR-STF decreases as the down-sampling factor increases. At the max down-sampling factor, where the HR frames are available only at the times when the first and last LR frames have been sampled, the PSNR of EDVR-STF is lower than that of EDVR in case 2 (34.31~dB vs 34.45~dB) and higher than that of EDVR in case 1 (32.34~dB vs 31.31~dB). However, EDVR-STF still outperforms all other algorithms in most situations considered, demonstrating the value of fusing HR video streams to boost the overall accuracy of the image sequence reconstruction. At each temporal down-sampling factor in both case 1 and case 2, both EDVR and EDVR-STF perform significantly better than bicubic interpolation. The other two performance indices show consistent patterns among the 4 methods and as the temporal down-sampling factor increases. In the supplementary information (Figure~\ref{fig:S2} and Figure~\ref{fig:S3}), the same results with larger frame separations (2 and 3) of the LR image sequence are shown. Consistent trends in all performance indices can be observed.

\subsection{Normalized attention scores}
\begin{figure}
    \label{fig:4}
    \hspace*{-1.5cm}
    \includegraphics[width=1.2\textwidth]{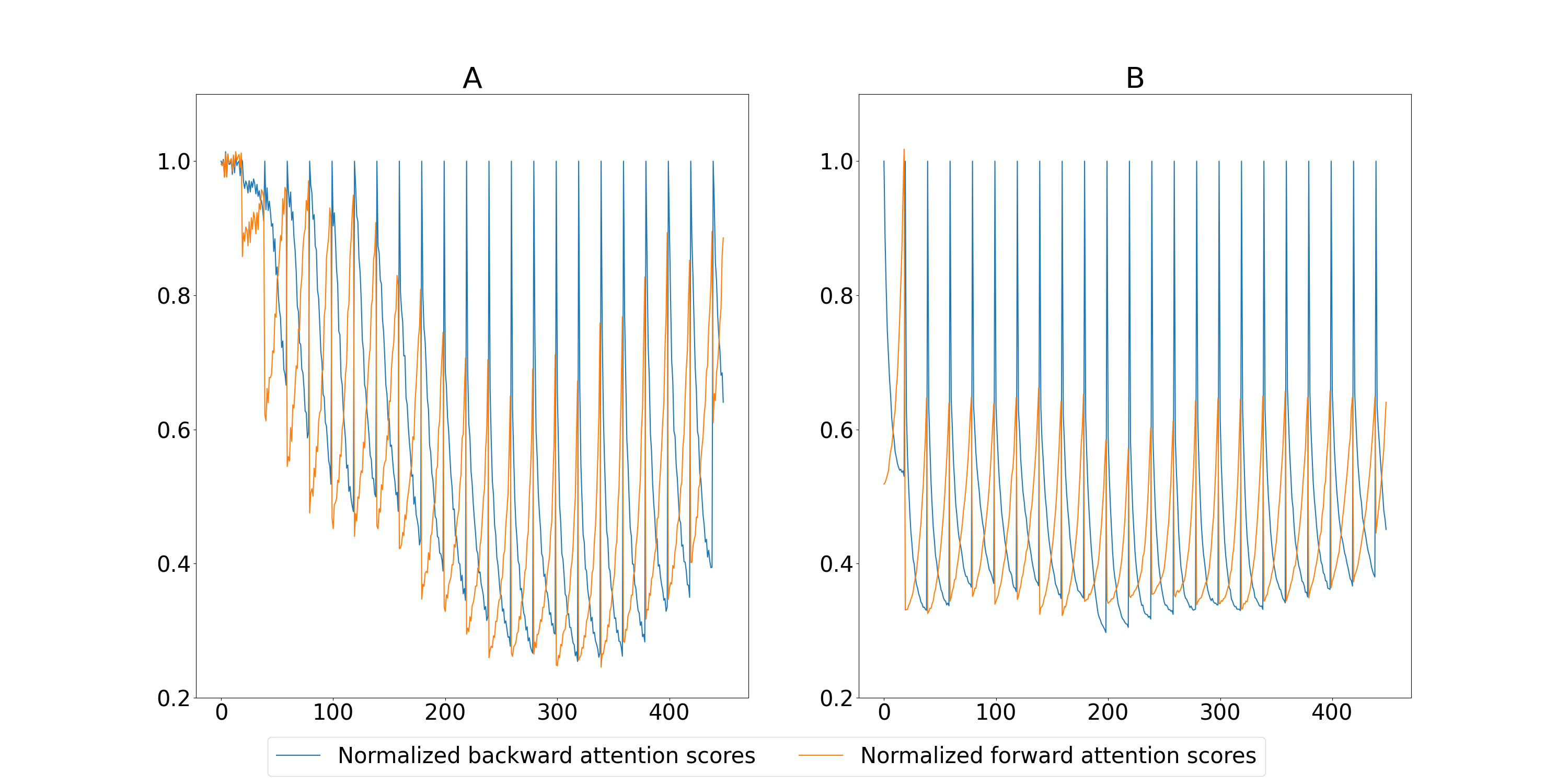}
    \caption{Normalized backward and forward attention scores of the EDVR-STF model to predict continuous target HR frames in case 1 and case 2. For both cases, the LR frame separation was set to 1 and the HR image sequence was down-sampled by a factor of 20. No Poisson noise was generated for the testing data.}
\end{figure}

\begin{figure}
    \label{fig:S4}
    \hspace*{-3cm}
    \includegraphics[width=1.5\textwidth]{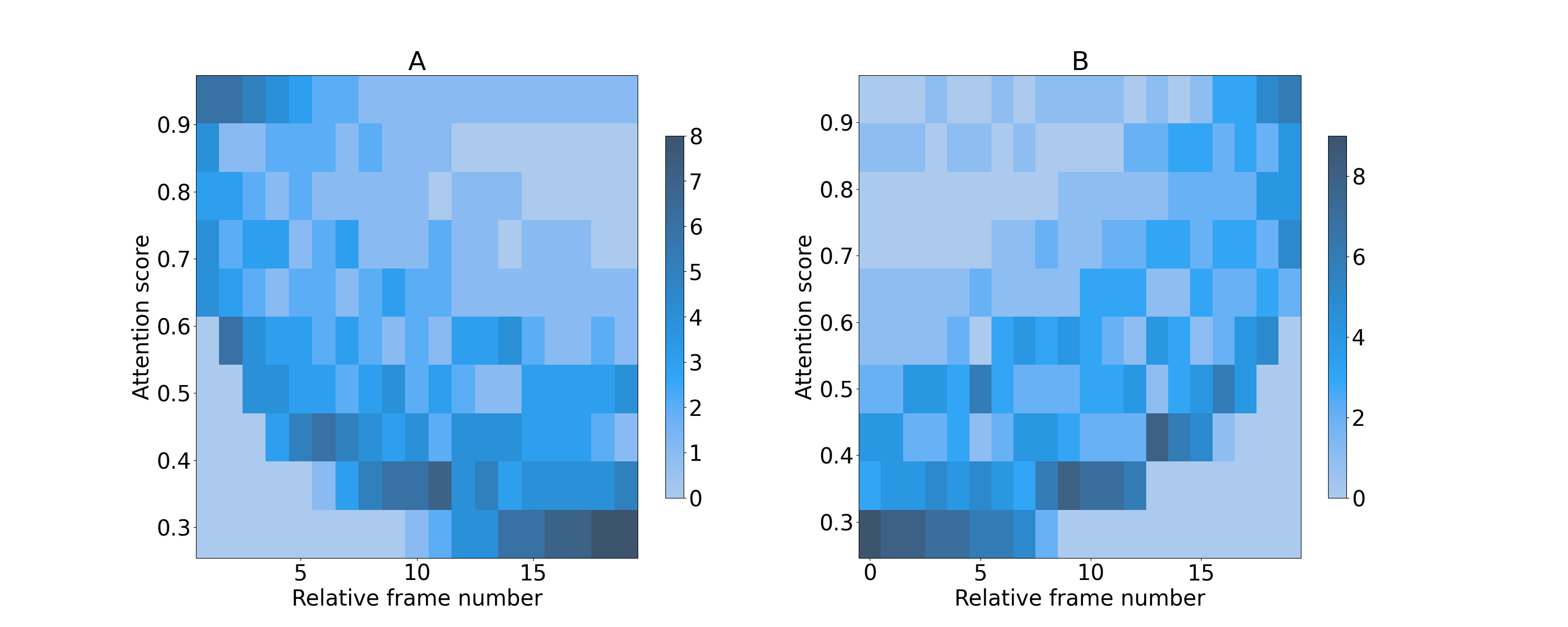}
    \caption{2-D histograms of the normalized backward attention scores (A) and forward attention scores (B) and the relative frame numbers when the EDVR-STF model was evaluated on continuous frames from one testing data set (case 1). In the illustrated case, the LR frame separation was set to 1 and the HR image sequence was down-sampled by a factor of 20. No Poisson noise was generated for the testing data.}
\end{figure}

Figure~\ref{fig:4} shows the normalized backward and forward attention scores obtained from the same 2 testing data sets. Between each pair of subsequent HR frames, the backward attention scores decrease as the target HR frame is far away from its preceding HR frame, and the forward attention scores increase accordingly. Figure~\ref{fig:S4} further shows the temporal distributions of the normalized backward and forward attention scores in case 1 as the frame number of the target HR frame changes in relation to its neighbouring HR frames used for the reconstruction. In Figure~\ref{fig:S4}-A when the target frame is close to its preceding HR frame (e.g., 1 frame after it), the corresponding attention scores are distributed primarily above 0.8. As the target frame is away from its preceding HR frame, a decreasing trend can be observed in the backward attention scores, which signifies the less important role the preceding HR frame plays in improving the resolution in the target frame. When the target frame number is close to that of its succeeding HR frame (e.g., 19 frames after the preceding frame), the majority of the attention scores are below 0.3. The opposite trend can be observed in the case of the forward attention scores as illustrated in Figure~\ref{fig:S4}-B.

\subsection{Computation times}

\begin{figure}
    \label{fig:9}
    \centering
    \includegraphics[width=0.75\textwidth]{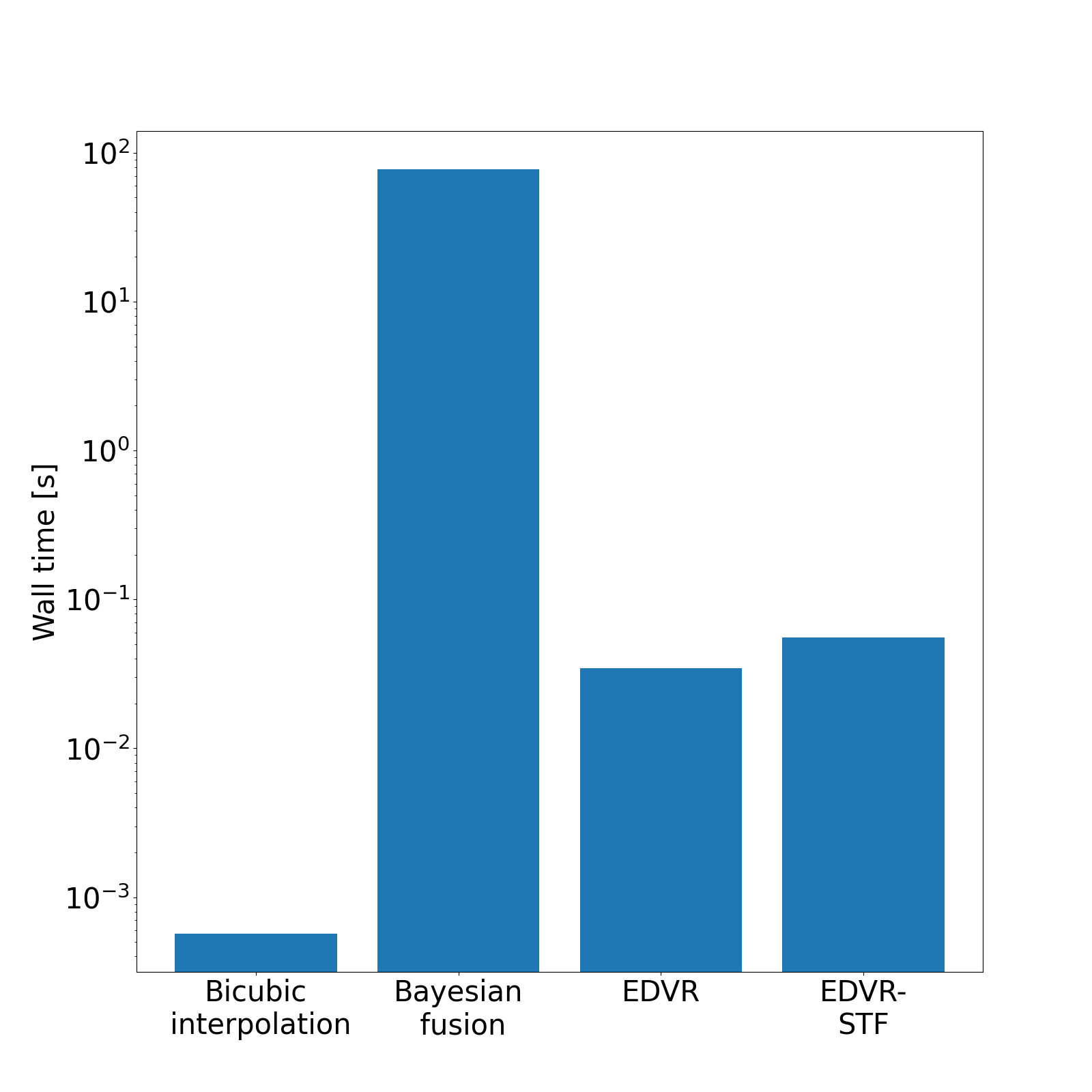}
    \caption{Wall times of the bicubic interpolation, Bayesian fusion framework, EDVR, and EDVR-STF reported as the median over all 100 realizations.}
\end{figure}

Computation times of the 4 algorithms are shown in Figure~\ref{fig:9}. The bicubic interpolation and Bayesian fusion framework both execute end-to-end, with a wall time of 0.0006~s and 77.4044~s, respectively, on an Intel Xeon Gold 6334 CPU@ 3.60 GHz to reconstruct an HR image with 400×1024 pixels. The EDVR and EDVR-STF models require a fixed time for model training. Once the models are trained, inference executes end-to-end with a wall time of 0.0408~s and 0.0665~s, respectively, on 1 Nvidia A-100 SXM4 GPU (40 GB memory) to reconstruct an HR image with 400×1024 pixels. The bicubic interpolation was based on the \cite{OpenCV} function resize. The Bayesian fusion algorithm was custom built following details of \cite{xue2017bayesian}. The wall time of each method was estimated as the median wall time over 100 runs.

\section{Discussion}

In this paper, we present the first study that attempts to fuse two x-ray image sequences that are optimized for imaging speed and spatial resolution, respectively, and reconstruct the target image sequence that enjoys both high spatial resolution and high frame rate. In the current workflow of an UHS x-ray experiment, each individual high-speed camera is optimized to fulfil a dedicated role. However, given its optimal configuration, one single camera's utility usually only covers a partial aspect of the entire envelope of the high-speed x-ray radiography experiment. In order to continually increase the scientific value that can be realized by the imaging experiment, the synergistical effect of multiplexing high-speed cameras must be better exploited. In this respect, a model that coordinates functionally distinct cameras and consolidates their respective imaging capabilities could greatly propel the development of the entire user community of UHS x-ray radiography. Due to the excellent performance of machine learning in modeling complex data structures as in the application of spatio-temporal analysis \cite{shi}, we concentrated on a deep convolutional neural network in this study, termed ``EDVR-STF''.

In order to benchmark the proposed EDVR-STF model, we compared its performance with three other methods, namely, the baseline EDVR model, the Bayesian fusion framework, and the single frame bicubic interpolation, based on three commonly used metrics, i.e., PSNR, AAD, and SSIM. On one hand, the baseline EDVR model used the same LR images as input and was trained using the same configuration, differing from EDVR-STF only in the feature extraction branch to couple the HR feature maps. The purpose of including the baseline EDVR model was to understand the effect of mixing HR frames with LR frames on restoring the target HR frame. On the other hand, the Bayesian fusion framework used the same HR and LR images as in EDVR-STF as input and was intended to provide a benchmark on the effect of utilizing a deep learning-based spatio-temporal fusion model.

It should be noted that in its current implementation, the Bayesian framework only learns pixel-level calibration models, which is in drastic contrast to the deep learning-based framework. When the 4 image reconstruction approaches were compared, two variables were chosen, namely, the frame separations of the HR image sequence, and the Poisson noise level in the LR image sequence. In the evaluation of high-speed cameras, the frame rate is one of the most crucial specifications of the equipment, which could be conveniently used to stratify the instruments according to their distinct use scenarios. In this study, we concentrated on varying the frame separation of the HR image sequence, whereas the LR image sequence was intended to provide a well sampled time series of the target event. As a result, the configuration of the LR image sequence was contextualized in the specific scientific community utilizing the x-ray experiment. Poisson noise has been widely recognized as the dominant source of image noise under the majority of illumination conditions \cite{Hasinoff}. In the x-ray experiment involving indirect detectors, the Poisson noise occurs with the random arrival of photons that get converted from the incoming x-ray, also termed as the “shot noise”. Due to its signal-dependent nature, the shot noise often varies with the physical configurations of an imaging system, such as the exposure levels and sensor gains \cite{Hasinoff}. In practice, physical configurations of an ultra-high-speed cameras have often resulted in higher noise levels than the high-speed cameras \cite{Ren}. Therefore, in this study, Poisson noise was generated only in the LR images to demonstrate method robustness to the shot noise as a proof of concept. Restoration of video quality in the presence of more complex types of process noise during the x-ray imaging experiment will be left in the future to investigate. Testing image reconstruction algorithms on these configurations could thus allow efficient assessment of the feasibility of novel experimental protocols for high-speed x-ray radiography.

All reconstruction algorithms were tested on two independent datasets. From both datasets, EDVR-STF showed an overall higher performance than that of EDVR and the Bayesian fusion framework, under each frame separation condition, which in turn demonstrates the value of coupling neighbouring HR image features with LR image features and of using a deep learning-based image reconstruction approach. On the other hand, the bicubic interpolation showed the lowest performance. Since the bicubic interpolation did not use any other frames than the LR reference frame itself, the results indicate effective data association of all other methods.

To allow a clear understanding of the performance of the proposed EDVR-STF model, we quantified the utilization of each of the two HR images as the input to reconstruct the target HR image. In particular, we derived numerical characteristics, termed as the ``forward attention score'' and ``backward attention score'' from the attention map between each of the HR image's feature maps and the reference LR feature maps. To reconstruct different target HR frames in an image sequence, the same set of input HR frames is not likely to play the same role. As a result, availability of quantitative metrics could provide valuable insight into the model performance under different device settings and with complex temporal dynamics of the underlying event. Since feature fusion is only one module of the entire spatio-temporal fusion model, interpretation of abnormally low attention scores needs more caution toward complex confounding effects. Nonetheless, the attention scores can be used to indicate the internal states of the model at inference time, and, along with the context information of the specific x-ray imaging experiments, identify unwanted working conditions of the equipment.

The proposed EDVR-STF model was intended to show preliminary results of applying a deep learning-based spatio-temporal fusion framework to solve the task of HR image sequence reconstruction. As a result, we concentrated on the specific configuration of fusing HR and LR image sequences with a fixed 4 times difference in each spatial dimension while allowing the model performance to scale with variable frame rates that can be configured for the HR video. As the primary challenge of high-speed x-ray radiography is in the large difference in the camera frame rate, we expect the proposed model architecture to be able to handle the majority of scenarios from the user community. In practice, the proposed model could be combined with other super resolution/spatio-temporal fusion methods to enable variable spatial up-sampling factors around the fixed one. The proposed model training and inference could also be extended to support other tasks in a light source, such as improving the acquisition and reconstruction workflow of the tomographic imaging data \cite{liu2019deep,liu2020tomogan,benmore2022advancing}. In this work, we pretrained the model on a diverse training data set outside the domain of x-ray imaging, and transferred the model to a dedicated x-ray image set from the domain of additive manufacturing. The effort to further improve model generalizability by curating a diverse training dataset within the x-ray imaging modality, from an authentic dual camera x-ray imaging setting, improving the model architecture hence scalability with increasing training sample size, and creating more general imaging conditions at training time will be left to be investigated in the future, with a foundation model trained on multiple image restoration tasks \cite{ma2024pretraining}.

\section{Conclusions}
In this paper, we investigated the use of a deep learning-based spatio-temporal fusion algorithm to integrate two image sequences with high frame rate and high spatial resolution, respectively, and reconstruct target image sequences with high frame rate, high spatial resolution, and high fidelity at the same time. The algorithm is implemented in Python and publicly available at 
\renewcommand\UrlFont{\color{blue}\rmfamily} \url{https://github.com/xray-imaging/XFusion}. With input image sequences of 4 times lower spatial resolution and 20 times lower frame rate, respectively, it achieved an average PSNR of more than 35 dB based on our test set with interpretable performance metrics of the attention scores. We therefore conclude that the proposed framework could reconstruct the target x-ray image sequence with high fidelity, with realistic physical configurations of the high-speed and ultra-high-speed cameras.

\section{Acknowledgements}
This research used resources of the Advanced Photon Source, a U.S. Department of Energy (DOE) Office of Science user facility and is based on work supported by Laboratory Directed Research and Development (LDRD) funding from Argonne National Laboratory, provided by the Director, Office of Science, of the U.S. DOE under Contract No. DE-AC02-06CH11357. This research used resources of the Argonne Leadership Computing Facility, a U.S. Department of Energy (DOE) Office of Science user facility at Argonne National Laboratory and is based on research supported by the U.S. DOE Office of Science, Advanced Scientific Computing Research Program, under Contract No. DE-AC02-06CH11357. This research was also supported by the Laboratory Directed Research and Development (LDRD) funding 2024-0026 from Argonne National Laboratory. The authors also acknowledge Frank E. Pfefferkorn (University of Wisconsin-Madison) for allowing the use of the friction stir welding dataset for model testing.

\supplementarysection
\subsection{Method} \label{sec:supp_method}
\subsubsection{Model architecture} \label{sec:supp_model_arch}
\begin{figure}
    \label{fig:schematic}
    \centering
    \includegraphics[width=1\textwidth]{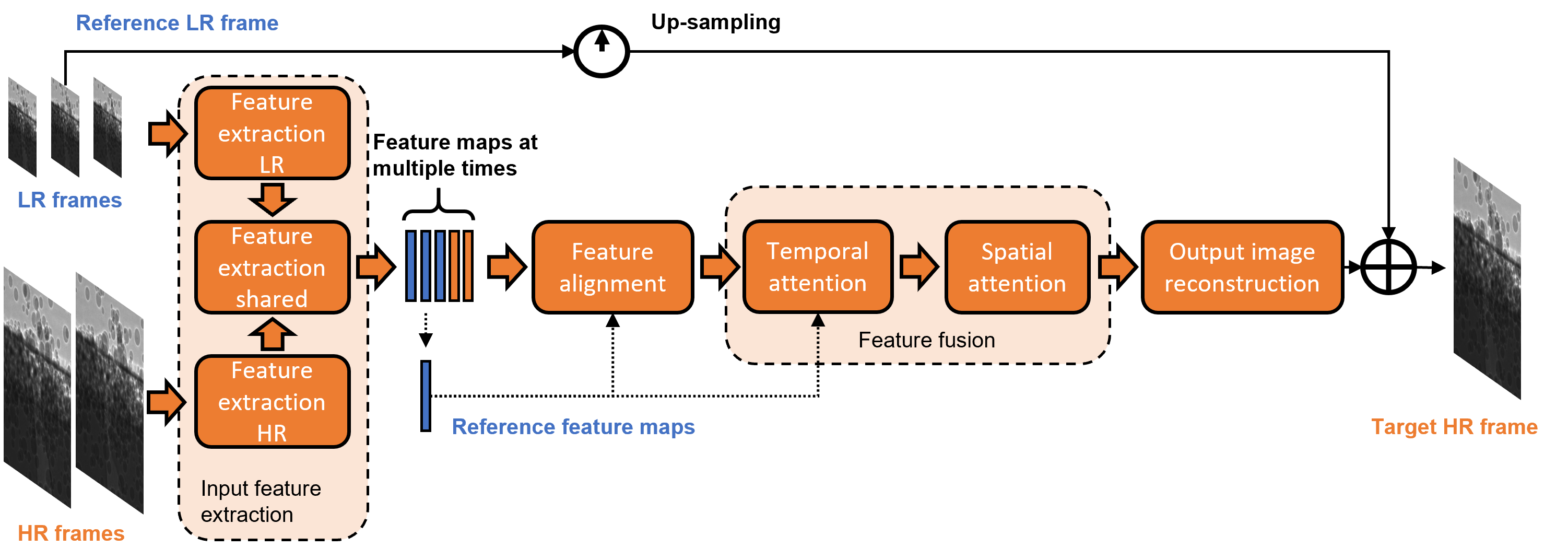}
    \caption{Architecture of the proposed spatio-temporal fusion model. Model inputs are consecutive LR and HR images that can be acquired from the corresponding high-speed and UHS cameras. Model output is a single HR image at the time when the reference LR image is acquired.}
\end{figure}

: as shown in Figure~\ref{fig:schematic}, the model consists of a cascade of four modules; namely, input feature extraction, feature alignment, feature fusion, and output image reconstruction. The input feature extraction module consists of independent and shared convolutional layers to process the LR and HR images separately and altogether. Each convolutional layer is followed by an activation layer to propagate extracted features between layers. 1 and 3 convolutional layers were included in the feature extraction modules for LR and HR images, respectively, and 5 residual blocks, each containing 2 convolutional layers with skip connections, were included in the subsequent shared feature extraction module. For the HR images, the last two convolution layers of their feature extraction module each was followed by an activation operation and a max pooling layer to reduce the resulting spatial dimensions of feature maps and keep them the same as those of the LR feature maps. A set of feature maps was thus output from the feature extraction module, corresponding to an input LR or HR frame at a distinct time point. The feature maps corresponding to the reference LR frame was identified as the ``reference feature maps''. The number of feature channels within each distinct set of feature maps was kept at 128 to balance the model complexity and the expressiveness of the learned feature maps.

The subsequent feature alignment, feature fusion, and output image reconstruction modules were configured according to the original EDVR model architecture \cite{Wang:19}. In particular, the feature alignment module used a pyramid of 3 levels, differing in the scale by a factor of 2 between subsequent levels in both spatial dimensions, to align each distinct set of feature maps with the reference feature maps in a coarse-to-fine manner. The feature fusion module consists of a temporal attention module followed by a spatial attention module to aggregate features unique to the HR images at the reference time. More specifically, the temporal attention module was used to weight each set of aligned feature maps on a pixel-by-pixel manner based on their similarity to the reference feature maps and blend them in the reference time. The spatial attention module further learned with a large receptive field affine transformations to be applied to each pixel from each channel of the temporally fused feature maps \cite{Tian}. The fused feature maps at the reference time were processed through the output image reconstruction module to approximate the residual image, which was added to the up-sampled reference LR image to restore the HR image at the same time point. In the reconstruction module, a cascade of 40 residual blocks, each containing 2 convolutional layers with skip connections were included, followed by 2 pixel shuffle up-sampling layers to recover the spatial dimension of the HR images \cite{Lim}.

\subsubsection{Model training}: a transfer learning strategy was adopted for model training to improve its generalization capability. In the pretraining stage, we used the realistic and diverse scenes (REDS) dataset of natural image sequences, which was an established benchmark on various video restoration tasks \cite{Nah}. REDS contained 240 training videos, 100 frames per video, and 720$\times$1280 pixels per frame. Each video was manually recorded to capture a diversity of objects and events. To reconcile the difference in the number of channels of the color images from REDS and the target x-ray images, we converted the color images to grayscale and established the pre-trained version of the EDVR-STF model that can be conveniently fine-tuned on x-ray data. In the fine-tuning stage, we used 547 videos recorded with a Photron FastCam SA-Z camera (Photron Inc., Japan) operated at a frame rate of 50 kHz during an operando high-speed synchrotron x-ray imaging experiment performed at the 32-ID beamline of the APS \cite{Ren}.

Four videos were randomly sampled from all the videos and held out as validation data and the remaining were used as the training data. Each video contained 500 frames with 400$\times$1024 pixels per frame. For the training data, random crops of 256$\times$256 and 64$\times$64 pixels were made on the corresponding HR and LR images to cover a spatially consistent region in each frame. Data augmentation included random horizontal flip, random vertical flip, and random rotation by 90 degrees, independently with a probability of 0.5. In addition, the frame separation between subsequent input LR frames was uniformly sampled at intervals 1, 2, and 3 at a time to increase the diversity of the input LR image sequence without significantly compromising the coherence among images. The frame separation between each of the two input HR images and the reference HR image was equal to that between subsequent input LR images scaled by a factor uniformly sampled from the discrete range of [-20,-1] (for the input HR image before the reference time) and [1,20] (for the input HR image after the reference time), respectively. The use of random frame separations among the input LR and HR was incorporated in an effort to desensitize the model from the underlying frame rate of the two video streams. Lastly, Poisson noise at each pixel of the LR images was simulated following \cite{Wu}. The Poisson noise model characterizes the fluctuation in the quantity of photons as incident on the detectors. In particular, the blank scan factor b0 was sampled in the range of 1 to 7 in the base-10 log scale (corresponding to a range of 10 to $10{\small,}000{\small,}000$ in the linear scale) and kept constant for each pair of 3 LR images to equalize the resulting low-dose images based on their PSNR levels (approximately 10~dB to 70~dB). The introduction of Poisson noise during model training could effectively improve model robustness to varying image qualities, increasing the flexibility of the dual camera system with more diverse acquisition settings at the individual modular level. The model was then trained to minimize the Charbonnier loss \cite{Wang:19} for a total of $300{\small,}000$ iterations with the Adam optimizer and an initial learning rate of 0.0001. The learning rate was updated following the cosine annealing scheduler. The model was trained using PyTorch framework and Argonne Leadership Computing Facility (ALCF) resources.

\subsubsection{Data pre-processing}\label{sec:supp_data_preproc}: for the first video type in the testing data, a total of 450 continuous frames were used, with 400$\times$1024 pixels per frame. For the second video type, a total of 450 continuous frames were used and each frame contained 1024$\times$1024 pixels. The original frames within each video were kept as the HR frames and binning \cite{peters2015precision} was applied to these frames to create another sequence of LR frames. For the resulting HR and LR image sequences, pixel values were clipped at the 0.35\textsuperscript{th} and 99.65\textsuperscript{th} percentiles according to their histograms, respectively, and the corresponding minimum and maximum pixel values across frames were in turn used to scale the pixel values of each individual frame following the min-max normalization rule. To further emulate different combinations of high-speed camera frame rates in a dual camera system, the frame separation between each LR image and its neighbouring LR images as inputs to the reconstruction algorithm was incremented from 1 to 3, and the original HR image sequence was resampled by keeping every 2\textsuperscript{nd} (2$\times$ lower frame rate), 10\textsuperscript{th} (10$\times$ lower frame rate), 20\textsuperscript{th} (20$\times$ lower frame rate) frames, and ultimately only the leading and trailing frames (450$\times$ lower frame rate). To further emulate different shot noises the two cameras were subjected to, 5 distinct levels of the blank scan factor $b_0$ for each x-ray video, leading to PSNRs (relative to the LR frames) of approximately 20~dB to 60~dB, (in increments of 10~dB), were used to synthesize Poisson noise (details refer to the previous paragraph) in the LR image sequence.

\subsubsection{Baseline reconstruction methods} \label{sec:supp_baselines}
\begin{figure}
    \label{fig:S1}
    \hspace*{-1.5 cm}
    \vspace*{-1.5 cm}
    \centering
    \includegraphics[width=1.2\textwidth]{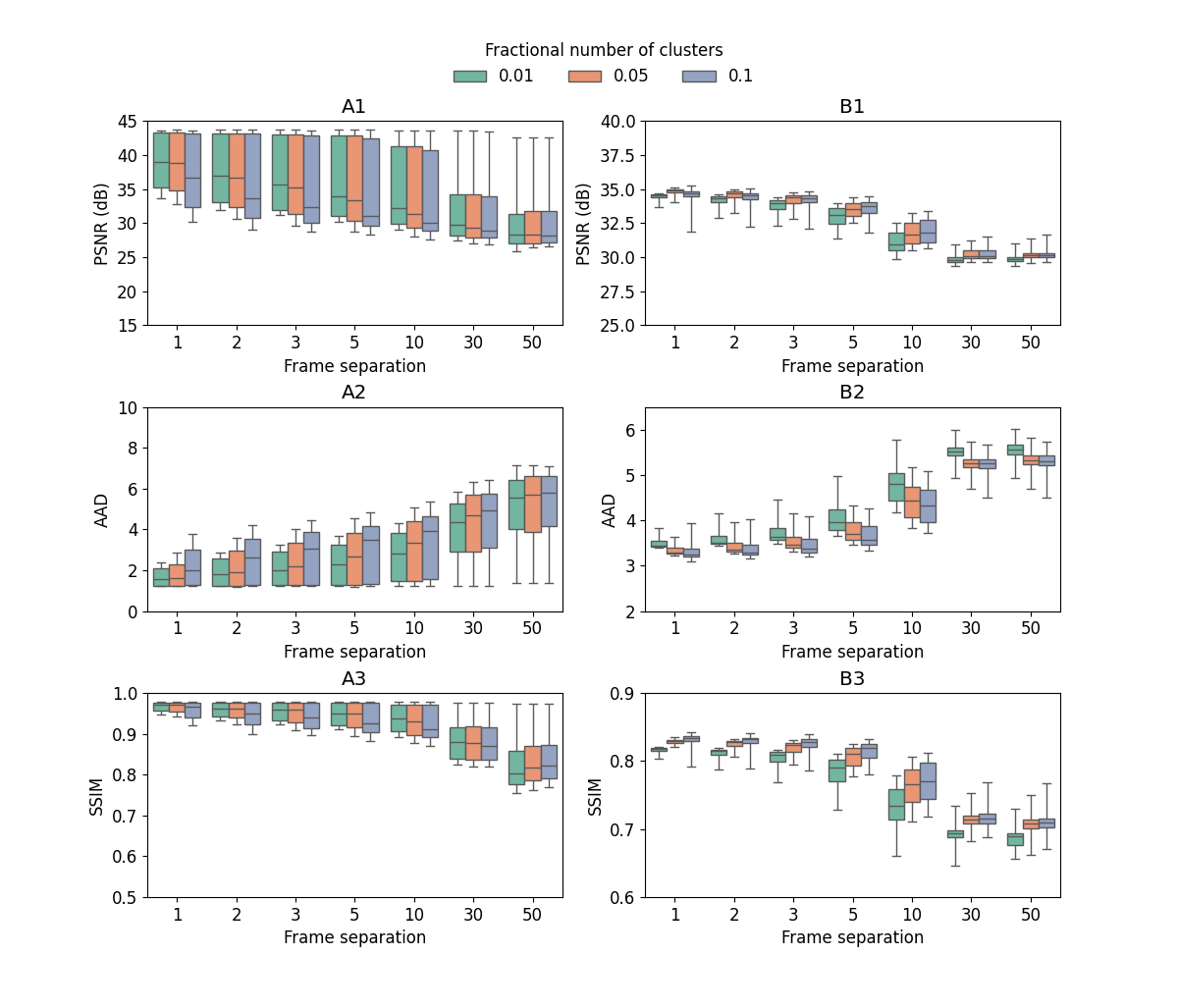}
    \caption{Reconstructed high resolution (HR) frame PSNR (dB) (row 1), AAD (row 2), and SSIM (row 3) each as a function of the cluster number fraction and the high/low resolution frame separation, based on Bayesian fusion. Results were evaluated on case 1 (column A) and case 2 (column B) of the main paper and presented as box plots. No Poisson noise was generated for the testing data.}
\end{figure}

: the Bayesian fusion framework used a similar input structure, i.e., 1 reference LR frame, 2 more LR frames and 2 more HR frames to estimate the HR frame at the same time as that of the reference LR frame. At every sampling time point, it first predicted the HR frame based on its temporal correlation with the two neighbouring HR frames and then updated it with the reference LR frame based on a generic image degradation model. By means of Gaussian models to represent the probabilistic change of pixel value across distinct spatial resolutions and in a time window, the proposed Bayesian framework led to a linear Kalman filter implementation to solve the overall fusion problem \cite{xue2017bayesian}. More specifically, the temporal change in each pixel value among subsequent frames was learned using a K-means clustering algorithm, with each cluster represented by a cluster mean vector and a covariance matrix and cluster assignment based on the Euclidean distance. Following \cite{xue2017bayesian}, each cluster's mean vector and covariance matrix were learned from the paired 3 LR frames across all positions sampled from the imaging plane. The number of clusters was determined as 5\% of the number of pixels in each dimension of the LR image and geometrically averaged across both the height and width dimensions, based on a preliminary analysis presented in (Figure~\ref{fig:S1}). For short time series as in our proposed input structure (i.e., 3 LR frames), the number of clusters is expected to be small. Our experiments suggested the use of $\sim$10~clusters. Although this choice of cluster number may slightly over partition the time series feature space under the Euclidean distance metric, it could provide additional flexibility when dealing with complex dynamics in the underlying event.
The image degradation model was a linear model with linear interpolation of HR pixel values from non-overlapping square-shaped kernels and additive Gaussian noise \cite{xue2017bayesian}. HR and LR frames before and after the reference LR frame were each paired to calibrate the degradation model. Since the Bayesian framework assumed the HR and LR frames to be obtained from the same time point, the corresponding frame separations were kept the same. For the baseline EDVR model, no HR frames were used during the inference, and the same 3 LR frames (one frame before, one frame at, and one frame after the reference time) were used to estimate the HR frame at the time of the reference LR frame. Analogous to the training procedure of EDVR-STF, it was pretrained on the same gray-scale REDS data set and then fine-tuned on the same x-ray images. The proposed EDVR-STF reconstruction algorithm was then tested, and the performance was compared with that of the bicubic interpolation, the Bayesian fusion and the baseline EDVR super resolution model.

\subsubsection{Attention scores}
\label{sec:supp_attention_scores}: 
From the temporal attention module of the EDVR-STF model, the attention map between each set of feature maps corresponding to each of the two distinct HR input frames and the reference LR feature maps was spatially averaged. The resulting attention score characterizes the importance of each of the HR frames in restoring the reference LR frame. The attention score between each LR frame and the preceding and succeeding HR frames from the corresponding image sequences, termed as ``backward attention score'' and ``forward attention score'', respectively, were then each grouped by the preceding HR frame and normalized by the attention score between the HR and LR frames both acquired at the time of the preceding HR frame. The corresponding frame indices were also offset by that of the same preceding HR frame. Bivariate distributions of the locally normalized attention scores and their frame indices relative to the preceding HR frame were obtained in the form of two 2-D histograms, one for the backward attention scores and one for the forward attention scores. The normalized backward and forward attention scores provide a means of assessing the utilization of the corresponding HR images by the EDVR-STF model, hence objective quality assessment of the fusion, in the absence of the ground truth HR images as required by conventional quality indices.

\begin{figure}
    \label{fig:S2}
    \hspace*{-1.5 cm}
    \vspace*{-1.5 cm}
    \includegraphics[width=1.2\textwidth]{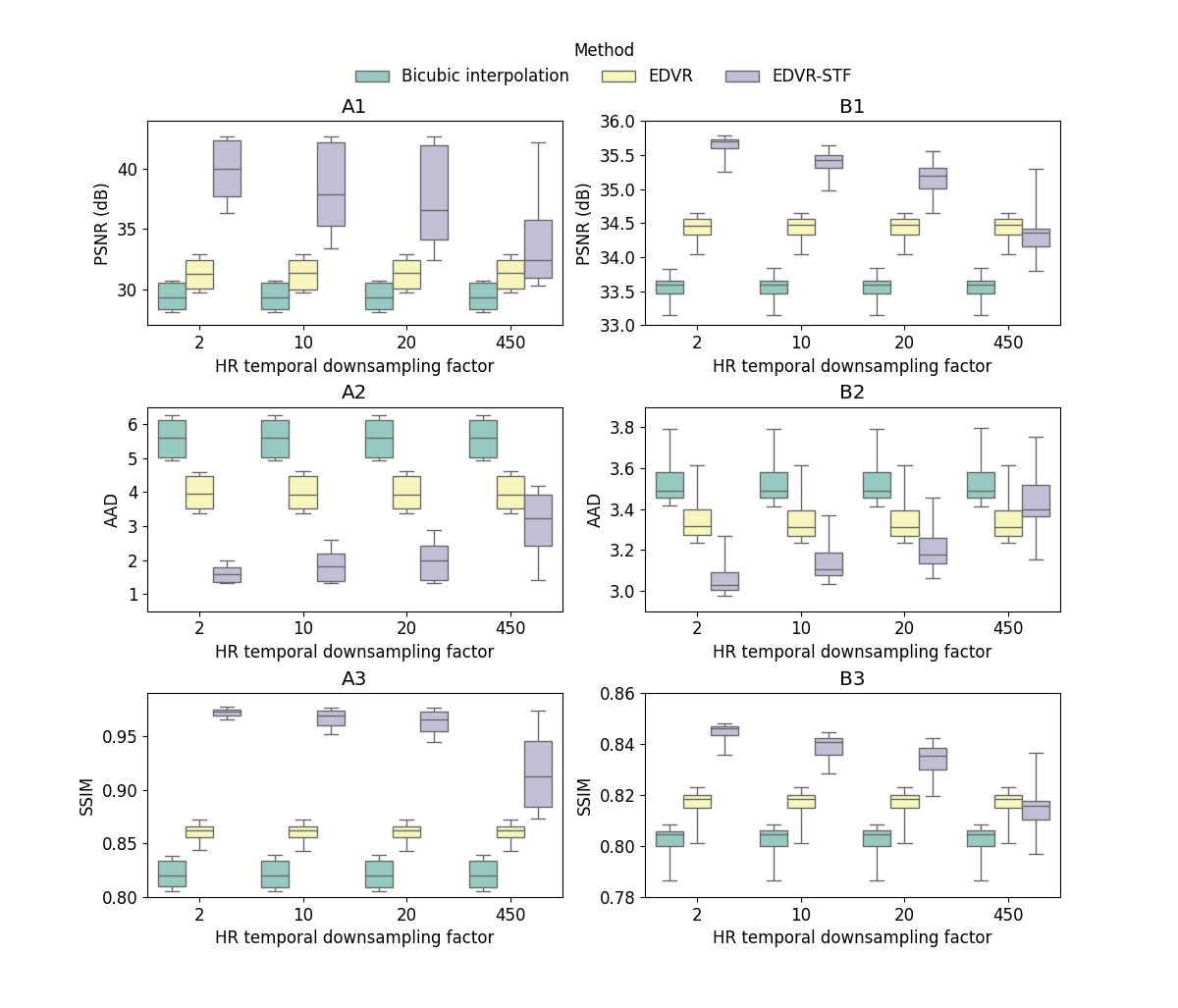}
    \caption{Reconstructed HR frame PSNR (dB) (row 1), AAD (row 2), and SSIM (row 3) each as a function of the HR frame sequence down-sampling factor, based on bicubic interpolation, EDVR, and EDVR-STF. Results were evaluated on case 1 (column A) and case 2 (column B) of the main paper and presented as box plots. LR separation was 2 across all plots. Under each testing condition, the same samples as used for EDVR-STF were used to test all other 2 algorithms. No Poisson noise was generated for the testing data.}
\end{figure}

\referencelist     


@article{Agiwal,
title = {Material flow visualization during friction stir welding using high-speed X-ray imaging},
journal = {Manufacturing Letters},
volume = {34},
pages = {62-66},
year = {2022},
issn = {2213-8463},
doi = {https://doi.org/10.1016/j.mfglet.2022.08.016},
url = {https://www.sciencedirect.com/science/article/pii/S2213846322001894},
author = {Hemant Agiwal and Mohammad {Ali Ansari} and Daniel Franke and Patrick Faue and Samuel J. Clark and Kamel Fezzaa and Shiva Rudraraju and Michael Zinn and Frank E. Pfefferkorn}
}

@ARTICLE{Chen,
  author={Chen, Lu and Ye, Mao and Ji, Luping and Li, Shuai and Guo, Hongwei},
  journal={IEEE Transactions on Broadcasting}, 
  title={Multi-Reference-Based Cross-Scale Feature Fusion for Compressed Video Super Resolution}, 
  year={2024},
  volume={},
  number={},
  pages={1-14},
  doi={10.1109/TBC.2024.3407517}}

@article{Douissard,
doi = {10.1088/1748-0221/7/09/P09016},
url = {https://dx.doi.org/10.1088/1748-0221/7/09/P09016},
year = {2012},
month = {sep},
publisher = {},
volume = {7},
number = {09},
pages = {P09016},
author = {P -A Douissard and  A Cecilia and  X Rochet and  X Chapel and  T Martin and  T van de Kamp and  L Helfen and  T Baumbach and  L Luquot and  X Xiao and  J Meinhardt and  A Rack},
title = {A versatile indirect detector design for hard X-ray microimaging},
journal = {Journal of Instrumentation}
}

@Article{Escauriza,
author={Escauriza, E. M.
and Duarte, J. P.
and Chapman, D. J.
and Rutherford, M. E.
and Farbaniec, L.
and Jonsson, J. C.
and Smith, L. C.
and Olbinado, M. P.
and Skidmore, J.
and Foster, P.
and Ringrose, T.
and Rack, A.
and Eakins, D. E.},
title={Collapse dynamics of spherical cavities in a solid under shock loading},
journal={Scientific Reports},
year={2020},
month={May},
day={21},
volume={10},
number={1},
pages={8455},
issn={2045-2322},
doi={10.1038/s41598-020-64669-y},
url={https://doi.org/10.1038/s41598-020-64669-y}
}

@BOOK{Hasinoff,
  title     = "Computer vision",
  editor    = "Ikeuchi, Katsushi",
  publisher = "Springer Nature",
  series    = "Computer Vision",
  edition   =  2,
  month     =  sep,
  year      =  2021,
  address   = "Cham, Switzerland",
  language  = "en"
}

@article{He,
title = {Multimodal fusion-based high-fidelity compressed ultrafast photography},
journal = {Optics and Lasers in Engineering},
volume = {181},
pages = {108363},
year = {2024},
issn = {0143-8166},
doi = {https://doi.org/10.1016/j.optlaseng.2024.108363},
url = {https://www.sciencedirect.com/science/article/pii/S0143816624003415},
author = {Yu He and Yunhua Yao and Yilin He and Chengzhi Jin and Zhengqi Huang and Mengdi Guo and Jiali Yao and Dalong Qi and Yuecheng Shen and Lianzhong Deng and Zhiyong Wang and Wei Zhao and Jinshou Tian and Yanhua Xue and Duan Luo and Zhenrong Sun and Shian Zhang}
}

@article{Koch,
title = {Lens coupled scintillating screen-CCD X-ray area detector with a high detective quantum efficiency},
journal = {Nuclear Instruments and Methods in Physics Research Section A: Accelerators, Spectrometers, Detectors and Associated Equipment},
volume = {348},
number = {2},
pages = {654-658},
year = {1994},
issn = {0168-9002},
doi = {https://doi.org/10.1016/0168-9002(94)90818-4},
url = {https://www.sciencedirect.com/science/article/pii/0168900294908184},
author = {Andreas Koch}
}

@article{Kornienko,
author = {Vassily Kornienko and David Andersson and Mehdi Stiti and Jonas Ravelid and Simon Ek and Andreas Ehn and Edouard Berrocal and Elias Kristensson},
journal = {Photon. Res.},
number = {7},
pages = {1712--1722},
publisher = {Optica Publishing Group},
title = {Simultaneous multiple time scale imaging for kHz\&\#x2013;MHz high-speed accelerometry},
volume = {10},
month = {Jul},
year = {2022},
url = {https://opg.optica.org/prj/abstract.cfm?URI=prj-10-7-1712},
doi = {10.1364/PRJ.451108},
}

@ARTICLE{Li,
  author={Li, Daoyu and Bian, Liheng and Zhang, Jun},
  journal={IEEE Journal of Selected Topics in Signal Processing}, 
  title={High-Speed Large-Scale Imaging Using Frame Decomposition From Intrinsic Multiplexing of Motion}, 
  year={2022},
  volume={16},
  number={4},
  pages={700-712},
  doi={10.1109/JSTSP.2022.3164524}}

@INPROCEEDINGS{Lim,
  author={Lim, Bee and Son, Sanghyun and Kim, Heewon and Nah, Seungjun and Lee, Kyoung Mu},
  booktitle={2017 IEEE Conference on Computer Vision and Pattern Recognition Workshops (CVPRW)}, 
  title={Enhanced Deep Residual Networks for Single Image Super-Resolution}, 
  year={2017},
  volume={},
  number={},
  pages={1132-1140},
  doi={10.1109/CVPRW.2017.151}}

@Article{Liu,
AUTHOR = {Hong, D. and Yao, J. and Wu, X. and Chanussot, J. and Zhu, X.},
TITLE = {SPATIAL-SPECTRAL MANIFOLD EMBEDDING OF HYPERSPECTRAL DATA},
JOURNAL = {The International Archives of the Photogrammetry, Remote Sensing and Spatial Information Sciences},
VOLUME = {XLIII-B3-2020},
YEAR = {2020},
PAGES = {423--428},
URL = {https://isprs-archives.copernicus.org/articles/XLIII-B3-2020/423/2020/},
DOI = {10.5194/isprs-archives-XLIII-B3-2020-423-2020}
}

@ARTICLE{Lu,
  author={Lu, Ming and Chen, Tong and Dai, Zhenyu and Wang, Dong and Ding, Dandan and Ma, Zhan},
  journal={IEEE Transactions on Multimedia}, 
  title={Decoder-Side Cross Resolution Synthesis for Video Compression Enhancement}, 
  year={2023},
  volume={25},
  number={},
  pages={2097-2110},
  doi={10.1109/TMM.2022.3142414}}

@article{Manin,
author = {Julien Manin and Scott A. Skeen and Lyle M. Pickett},
title = {{Performance comparison of state-of-the-art high-speed video cameras for scientific applications}},
volume = {57},
journal = {Optical Engineering},
number = {12},
publisher = {SPIE},
pages = {124105},
year = {2018},
doi = {10.1117/1.OE.57.12.124105},
URL = {https://doi.org/10.1117/1.OE.57.12.124105}
}

@inproceedings{Miyauchi,
author = {K. Miyauchi and Tohru Takeda and K. Hanzawa and Y. Tochigi and S. Sakai and R. Kuroda and H. Tominaga and R. Hirose and K. Takubo and Y. Kondo and S. Sugawa},
title = {{Pixel structure with 10 nsec fully charge transfer time for the 20m frame per second burst CMOS image sensor}},
volume = {9022},
booktitle = {Image Sensors and Imaging Systems 2014},
editor = {Ralf Widenhorn and Antoine Dupret},
organization = {International Society for Optics and Photonics},
publisher = {SPIE},
pages = {902203},
year = {2014},
doi = {10.1117/12.2042373},
URL = {https://doi.org/10.1117/12.2042373}
}

@InProceedings{Nah,
author = {Nah, Seungjun and Baik, Sungyong and Hong, Seokil and Moon, Gyeongsik and Son, Sanghyun and Timofte, Radu and Mu Lee, Kyoung},
title = {NTIRE 2019 Challenge on Video Deblurring and Super-Resolution: Dataset and Study},
booktitle = {Proceedings of the IEEE/CVF Conference on Computer Vision and Pattern Recognition (CVPR) Workshops},
month = {June},
year = {2019}
}

@article{Nguyen,
  title={Imaging with an ultra-high-speed video camera operating at 20 Mfps for 300 kpixels},
  author={Hoang Dung Nguyen and Tomoo Okinaka and Yasuhide Takano and Kohsei Takehara and Vu Truong Son Dao and Takeharu Goji Etoh},
  journal={Mechanical Engineering Journal},
  volume={3},
  number={6},
  pages={16-00286-16-00286},
  year={2016},
  doi={10.1299/mej.16-00286}
}

@article{etoh2017theoretical,
  title={The theoretical highest frame rate of silicon image sensors},
  author={Etoh, Takeharu Goji and Nguyen, Anh Quang and Kamakura, Yoshinari and Shimonomura, Kazuhiro and Le, Thi Yen and Mori, Nobuya},
  journal={Sensors},
  volume={17},
  number={3},
  pages={483},
  year={2017},
  publisher={MDPI}
}

@article{bonse1996x,
  title={X-ray computed microtomography ($\mu$CT) using synchrotron radiation (SR)},
  author={Bonse, Ulrich and Busch, Frank},
  journal={Progress in biophysics and molecular biology},
  volume={65},
  number={1-2},
  pages={133--169},
  year={1996},
  publisher={Elsevier}
}

@article{hartmann1975high,
  title={High-resolution direct-display x-ray topography},
  author={Hartmann, W and Markewitz, G and Rettenmaier, U and Queisser, HJ},
  journal={Applied Physics Letters},
  volume={27},
  number={5},
  pages={308--309},
  year={1975},
  publisher={American Institute of Physics}
}

@article{Ren,
author = {Zhongshu Ren  and Lin Gao  and Samuel J. Clark  and Kamel Fezzaa  and Pavel Shevchenko  and Ann Choi  and Wes Everhart  and Anthony D. Rollett  and Lianyi Chen  and Tao Sun },
title = {Machine learning–aided real-time detection of keyhole pore generation in laser powder bed fusion},
journal = {Science},
volume = {379},
number = {6627},
pages = {89-94},
year = {2023},
doi = {10.1126/science.add4667},
URL = {https://www.science.org/doi/abs/10.1126/science.add4667},
eprint = {https://www.science.org/doi/pdf/10.1126/science.add4667}}

@misc{shi,
      title={Machine Learning for Spatiotemporal Sequence Forecasting: A Survey}, 
      author={Xingjian Shi and Dit-Yan Yeung},
      year={2018},
      eprint={1808.06865},
      archivePrefix={arXiv},
      primaryClass={cs.LG},
      url={https://arxiv.org/abs/1808.06865}, 
}

@InProceedings{Tian,
author = {Tian, Yapeng and Zhang, Yulun and Fu, Yun and Xu, Chenliang},
title = {TDAN: Temporally-Deformable Alignment Network for Video Super-Resolution},
booktitle = {Proceedings of the IEEE/CVF Conference on Computer Vision and Pattern Recognition (CVPR)},
month = {June},
year = {2020}
}

@article{Wang:19,
  author       = {Xintao Wang and
                  Kelvin C. K. Chan and
                  Ke Yu and
                  Chao Dong and
                  Chen Change Loy},
  title        = {{EDVR:} Video Restoration with Enhanced Deformable Convolutional Networks},
  journal      = {CoRR},
  volume       = {abs/1905.02716},
  year         = {2019},
  url          = {http://arxiv.org/abs/1905.02716},
  eprinttype    = {arXiv},
  eprint       = {1905.02716},
  bibsource    = {dblp computer science bibliography, https://dblp.org}
}

@ARTICLE{Wang:04,
  author={Zhou Wang and Bovik, A.C. and Sheikh, H.R. and Simoncelli, E.P.},
  journal={IEEE Transactions on Image Processing}, 
  title={Image quality assessment: from error visibility to structural similarity}, 
  year={2004},
  volume={13},
  number={4},
  pages={600-612},
  doi={10.1109/TIP.2003.819861}}

@INPROCEEDINGS{Wu,
  author={Wu, Ziling and Bicer, Tekin and Liu, Zhengchun and De Andrade, Vincent and Zhu, Yunhui and Foster, Ian T.},
  booktitle={2020 IEEE/ACM Workshop on Machine Learning in High Performance Computing Environments (MLHPC) and Workshop on Artificial Intelligence and Machine Learning for Scientific Applications (AI4S)}, 
  title={Deep Learning-based Low-dose Tomography Reconstruction with Hybrid-dose Measurements}, 
  year={2020},
  volume={},
  number={},
  pages={88-95},
  doi={10.1109/MLHPCAI4S51975.2020.00017}}

@ARTICLE{Xiao,
  author={Xiao, Yi and Yuan, Qiangqiang and Jiang, Kui and He, Jiang and Lin, Chia-Wen and Zhang, Liangpei},
  journal={IEEE Transactions on Image Processing}, 
  title={TTST: A Top-k Token Selective Transformer for Remote Sensing Image Super-Resolution}, 
  year={2024},
  volume={33},
  number={},
  pages={738-752},
  doi={10.1109/TIP.2023.3349004}}

@Article{xue2017bayesian,
AUTHOR = {Xue, Jie and Leung, Yee and Fung, Tung},
TITLE = {A Bayesian Data Fusion Approach to Spatio-Temporal Fusion of Remotely Sensed Images},
JOURNAL = {Remote Sensing},
VOLUME = {9},
YEAR = {2017},
NUMBER = {12},
ARTICLE-NUMBER = {1310},
URL = {https://www.mdpi.com/2072-4292/9/12/1310},
ISSN = {2072-4292},
DOI = {10.3390/rs9121310}
}

@article{Zhang,
  author       = {Hongyan Zhang and
                  Yiyao Song and
                  Chang Han and
                  Liangpei Zhang},
  title        = {Remote Sensing Image Spatiotemporal Fusion Using a Generative Adversarial
                  Network},
  journal      = {{IEEE} Trans. Geosci. Remote. Sens.},
  volume       = {59},
  number       = {5},
  pages        = {4273--4286},
  year         = {2021},
  url          = {https://doi.org/10.1109/TGRS.2020.3010530},
  doi          = {10.1109/TGRS.2020.3010530},
  bibsource    = {dblp computer science bibliography, https://dblp.org}
}

@inproceedings{sugawa2013ultra,
  title={Ultra-high speed video imaging technologies},
  author={Sugawa, Shigetoshi},
  booktitle={ITE Technical Report 37.19},
  pages={9--14},
  year={2013},
  organization={The Institute of Image Information and Television Engineers}
}

@article{tochigi2012global,
  title={A global-shutter CMOS image sensor with readout speed of 1-Tpixel/s burst and 780-Mpixel/s continuous},
  author={Tochigi, Yasuhisa and Hanzawa, Katsuhiko and Kato, Yuri and Kuroda, Rihito and Mutoh, Hideki and Hirose, Ryuta and Tominaga, Hideki and Takubo, Kenji and Kondo, Yasushi and Sugawa, Shigetoshi},
  journal={IEEE Journal of Solid-State Circuits},
  volume={48},
  number={1},
  pages={329--338},
  year={2012},
  publisher={IEEE}
}

@article{ma2024pretraining,
  title={Pretraining a foundation model for generalizable fluorescence microscopy-based image restoration},
  author={Ma, Chenxi and Tan, Weimin and He, Ruian and Yan, Bo},
  journal={Nature Methods},
  pages={1--10},
  year={2024},
  publisher={Nature Publishing Group US New York}
}

@article{luo2012gas,
  title={Gas gun shock experiments with single-pulse x-ray phase contrast imaging and diffraction at the Advanced Photon Source},
  author={Luo, SN and Jensen, BJ and Hooks, DE and Fezzaa, K and Ramos, KJ and Yeager, JD and Kwiatkowski, K and Shimada, T},
  journal={Review of Scientific Instruments},
  volume={83},
  number={7},
  year={2012},
  publisher={AIP Publishing}
}

@article{Ramos,
doi = {10.1088/1742-6596/500/14/142028},
url = {https://dx.doi.org/10.1088/1742-6596/500/14/142028},
year = {2014},
month = {may},
publisher = {},
volume = {500},
number = {14},
pages = {142028},
author = {K J Ramos and B J Jensen and A J Iverson and J D Yeager and C A Carlson and D S Montgomery and D G Thompson and K Fezzaa and D E Hooks},
title = {In situ investigation of the dynamic response of energetic materials using IMPULSE at the Advanced Photon Source},
journal = {Journal of Physics: Conference Series}
}

@article{Parab,
author = "Parab, Niranjan D. and Zhao, Cang and Cunningham, Ross and Escano, Luis I. and Fezzaa, Kamel and Everhart, Wes and Rollett, Anthony D. and Chen, Lianyi and Sun, Tao",
title = "{Ultrafast X-ray imaging of laser{--}metal additive manufacturing processes}",
journal = "Journal of Synchrotron Radiation",
year = "2018",
volume = "25",
number = "5",
pages = "1467--1477",
month = "Sep",
doi = {10.1107/S1600577518009554},
url = {https://doi.org/10.1107/S1600577518009554}
}

@online{OpenCV,
  author = {OpenCV},
  year = 2024,
  url = {https://opencv.org/},
  urldate = {2024-08-12}
}

@misc{basicsr,
  author =       {Xintao Wang and Liangbin Xie and Ke Yu and Kelvin C.K. Chan and Chen Change Loy and Chao Dong},
  title =        {{BasicSR}: Open Source Image and Video Restoration Toolbox},
  howpublished = {\url{https://github.com/XPixelGroup/BasicSR}},
  year =         {2022}
}

@inproceedings{peters2015precision,
  title={Precision of FLEET velocimetry using high-speed CMOS camera systems},
  author={Peters, Christopher J and Danehy, Paul M and Bathel, Brett F and Jiang, Naibo and Calvert, Nathan and Miles, Richard B},
  booktitle={31st AIAA Aerodynamic Measurement Technology and Ground Testing Conference},
  pages={2565},
  year={2015}
}

@article{xue2019video,
  title={Video enhancement with task-oriented flow},
  author={Xue, Tianfan and Chen, Baian and Wu, Jiajun and Wei, Donglai and Freeman, William T},
  journal={International Journal of Computer Vision},
  volume={127},
  pages={1106--1125},
  year={2019},
  publisher={Springer}
}

@inproceedings{werlberger2011optical,
  title={Optical flow guided TV-L 1 video interpolation and restoration},
  author={Werlberger, Manuel and Pock, Thomas and Unger, Markus and Bischof, Horst},
  booktitle={Energy Minimization Methods in Computer Vision and Pattern Recognition: 8th International Conference, EMMCVPR 2011, St. Petersburg, Russia, July 25-27, 2011. Proceedings 8},
  pages={273--286},
  year={2011},
  organization={Springer}
}

@article{fransens2007optical,
  title={Optical flow based super-resolution: A probabilistic approach},
  author={Fransens, Rik and Strecha, Christoph and Van Gool, Luc},
  journal={Computer vision and image understanding},
  volume={106},
  number={1},
  pages={106--115},
  year={2007},
  publisher={Elsevier}
}

@inproceedings{caballero2017real,
  title={Real-time video super-resolution with spatio-temporal networks and motion compensation},
  author={Caballero, Jose and Ledig, Christian and Aitken, Andrew and Acosta, Alejandro and Totz, Johannes and Wang, Zehan and Shi, Wenzhe},
  booktitle={Proceedings of the IEEE conference on computer vision and pattern recognition},
  pages={4778--4787},
  year={2017}
}

@article{wang2020deep,
  title={Deep video super-resolution using HR optical flow estimation},
  author={Wang, Longguang and Guo, Yulan and Liu, Li and Lin, Zaiping and Deng, Xinpu and An, Wei},
  journal={IEEE Transactions on Image Processing},
  volume={29},
  pages={4323--4336},
  year={2020},
  publisher={IEEE}
}

@inproceedings{dai2017deformable,
  title={Deformable convolutional networks},
  author={Dai, Jifeng and Qi, Haozhi and Xiong, Yuwen and Li, Yi and Zhang, Guodong and Hu, Han and Wei, Yichen},
  booktitle={Proceedings of the IEEE international conference on computer vision},
  pages={764--773},
  year={2017}
}

@inproceedings{zhu2019deformable,
  title={Deformable convnets v2: More deformable, better results},
  author={Zhu, Xizhou and Hu, Han and Lin, Stephen and Dai, Jifeng},
  booktitle={Proceedings of the IEEE/CVF conference on computer vision and pattern recognition},
  pages={9308--9316},
  year={2019}
}

@article{benmore2022advancing,
  title={Advancing AI/ML at the Advanced Photon Source},
  author={Benmore, Chris and Bicer, Tekin and Chan, Maria KY and Di, Zichao and G{\"u}rsoy, Dog˘ a and Hwang, Inhui and Kuklev, Nikita and Lin, Dergan and Liu, Zhengchun and Lobach, Ihar and others},
  journal={Synchrotron Radiation News},
  volume={35},
  number={4},
  pages={28--35},
  year={2022},
  publisher={Taylor \& Francis}
}

@inproceedings{liu2019deep,
  title={Deep learning accelerated light source experiments},
  author={Liu, Zhengchun and Bicer, Tekin and Kettimuthu, Rajkumar and Foster, Ian},
  booktitle={2019 IEEE/ACM Third Workshop on Deep Learning on Supercomputers (DLS)},
  pages={20--28},
  year={2019},
  organization={IEEE}
}

@article{liu2020tomogan,
  title={TomoGAN: low-dose synchrotron x-ray tomography with generative adversarial networks: discussion},
  author={Liu, Zhengchun and Bicer, Tekin and Kettimuthu, Rajkumar and Gursoy, Doga and De Carlo, Francesco and Foster, Ian},
  journal={JOSA A},
  volume={37},
  number={3},
  pages={422--434},
  year={2020},
  publisher={Optica Publishing Group}
}
\end{document}